 \documentclass[12pt]{article}
 \usepackage{graphicx}
 \usepackage{amssymb}
 \usepackage{amsmath}
 \usepackage{amsfonts}
 \usepackage{revsymb}

 \newcommand{\be}{\begin{equation}}
\newcommand{\ee}{\end{equation}}
\newcommand{\bea}{\begin{eqnarray}}
\newcommand{\eea}{\end{eqnarray}}
\newcommand{\bean}{\begin{eqnarray*}}
\newcommand{\eean}{\end{eqnarray*}}

\def\LL{{\cal L}}
\def\TT{{\cal T}}

\def\VV{{\cal V}}
\def\p{{\partial}}
\def\ov{\over}
\def\apr{{\alpha'}}

\begin{document}


\begin{titlepage}

\title{
\hfill\parbox{4cm}
{\normalsize {\tt USTC-ICTS-08-02}}\\
\vspace{1cm}\bf \Large High spin baryon in hot strongly coupled
plasma}

\author{\bf{\small Miao
Li}$\,^{a,b}$ \footnote{E-mail: mli@itp.ac.cn} $\ $, $\ $ \bf{\small
Yang Zhou}$\,^{a,b}$\footnote{E-mail: yzhou@itp.ac.cn} and
\bf{\small Push Pu}$\,^{c}$\footnote{E-mail: sldvm@mail.ustc.edu.cn}
\vspace{5mm}\\[12pt]\small
$^a$~{\it Interdisciplinary Center of Theoretical Studies,University of Science}\\
\small{\it and Technology of China, Anhui 230026, China}\\
\small$^b$~{\it Institute of Theoretical Physics,Academia Sinica, Beijing 100190, China}\\
\small$^c$~{\it Department of Modern Physics, University of Science and Technology of China,}\\
\small{\it Auhui 230026, China}\\
\\
}

\date{}

\maketitle

\begin{abstract}

We consider a strings-junction holographic model of a probe baryon
in the finite-temperature supersymmetric Yang-Mills dual of the
AdS-Schwarzschild black hole background. In particular, we
investigate the screening length for a high spin baryon composed of
rotating $N_c$ heavy quarks. To rotate quarks by finite force, we
put a hard infrared cutoff in the bulk and give quarks a finite
mass. We find that $N_c$ microscopic strings are embedded reasonably
in the bulk geometry when they have finite angular velocity
$\omega$, similar to the meson case. By defining the screening
length as the critical separation of quarks, we compute the $\omega$
dependence of the baryon screening length numerically and obtain a
reasonable result which shows that baryons with high spin dissociate
more easily. Finally, we discuss the relation between $J$ and $E^2$
for baryons.
\end{abstract}
\end{titlepage}

\tableofcontents

\bigskip

\setcounter{equation}{0}
 \section{Introduction}

  Recently there has been much interest in studying strongly coupled QCD
in terms of AdS/CFT correspondence. The first example of AdS/CFT
correspondence is provided by the duality between the classical
gravity in AdS$_5$$\times$S$^5$ and $\mathcal {N}=4$ supersymmetric
Yang-Mills theory ~\cite{AdS/CFT}. Quarks can be introduced as
fundamental, light ones, which appear in the action. They also can
be introduced as external probes, which are often infinitely heavy
and not included in the action. In the first case we need introduce
the flavor branes, and a background geometry dual to supersymmetric
SU($N_c$) Yang-Mills theory with $N_f$ additional hypermultiplets
~\cite{Karch:2002}. In this paper we treat the heavy quarks as
probes. Since QCD is confining, we always consider a quark and an
antiquark together. To measure the interaction potential between
them, we choose a physical, gauge invariant object named Wilson loop
\be
 W[C]=tr P\exp \int iA_{\mu} dx^{\mu}\;.
\ee By choosing the contour $C$ as a rectangle with length of
$\mathcal {T}$ in the time direction and two sides $L$ in a space
direction. When $L<<\mathcal {T}$, we can extract the potential
$V_{q\bar{q}}(L)$ by \be <W(C)>_0=\exp (-V_{q\bar{q}}(L)\mathcal
{T})\;. \ee In SYM at zero temperature, the potential between probe
quark and antiquark separated by $L$ is given by
~\cite{Maldacena:1998wl,Rey:1998wl}
 \be
V_{q\bar{q}}(L)=-\frac{4\pi^2}{\Gamma(1/4)^4}\frac{\sqrt{2g_{YM}^2N}}{L}
\ee valid in the large $N_c$ and large $\lambda$ limit. In AdS/CFT
framework, we obtain the potential by computing the action of a
hanging string with the infinite boundary $C$ of world sheet.

  The dual gravity background of SYM gauge theory at a nonzero
temperature is an AdS metric with a black hole. The potential
between two quarks then becomes~\cite{Rey:1998pl} \be
 V(L,T) \approx  \sqrt{\lambda} f (L,T) (L
 < L_c)\qquad
 V(L,T)\approx  \lambda^0 g(L,T) ( L > L_c),
\ee where $L_c$ is a critical separation, at which the qualitative
behavior of the interaction potential between a quark and an
antiquark changes. Roughly speaking, if $L<L_c$, a quark and an
antiquark form a bound state, while if $L>L_c$, they interact weakly
relatively. In a hot plasma, the critical length is related to the
temperature.

\subsection{Meson in strongly coupled gauge field}
  We consider a quark-antiquark bound state as a meson and define
$L_s=L_c$ as the screening length of meson , which is the critical
distance between the quark and antiquark. At separation of $L_s$ ,
the qualitative behavior of interaction potential changes. It is
also believed that dissociation happens when the quark-antiquark
separation is larger than $L_s$. For example, once the screening
length at a certain temperature is smaller than the radius of meson
(such as the $J/\psi$) in the quark gluon plasma, the quarkonium
will dissociate. The dissociation of meson can be used as a signal
of the formation of QGP as well as its temperature profile. $L_s$
can be expressed as a decreasing function of temperature and we
usually write $L_s=\alpha/T_{diss}$ in a hot plasma, where
$T_{diss}$ is the dissociation temperature, and $\alpha$ is a
constant determined by the characteristic QCD plasma. Lattice
calculations give more details of the static potential between a
heavy quark and antiquark in hot
QCD~\cite{Kaczmarek:2004,Kaczmarek:2005}. But the potential between
moving or rotating quarks is hard to calculate in lattice.

  In recent works~\cite{Liu:2006sw,Liu:2006wl,Chernicoff:2006hi}, the screening
length of meson is analyzed in the case of a moving quark-antiquark
pair. The velocity dependence of the screening length have been
calculated\footnote{It's pointed out in
work~\cite{Chernicoff:2006hi} that over the entire velocity range
$0\leq v\leq1$ the behavior of the screening length is actually
closer to $(1-v^2)^{1/3}$ than $(1-v^2)^{1/4}$ (this can be seen
most clearly in Fig.12 in a more recent
paper~\cite{Chernicoff:2008sa}), and it is only in the
ultra-relativistic regime $v\rightarrow 1$ that the behavior is in
accord with the exponent 1/4 determined analytically in
~\cite{Liu:2006sw}.}

\be\label{ls_v}
 L_s (v,T) \simeq L_s (0,T) (1-v^2)^{1/4}
 \propto \frac{1}{T_{diss}}(1-v^2)^{1/4}\ .
\ee From this equation, we see that the meson can not run too fast
in order to avoid dissociation. At a certain temperature, the
limiting velocity can be obtained by calculating the dispersion
relation of mesons which are described as fluctuation of the flavor
brane in the dual gravity background~\cite{Ejaz:2007,Mateos:2007vn}.

  By studying the spectral function of meson fluctuation on the
embedded flavor branes, we can obtain more information about the
phase diagram of a strongly coupled gauge theory. The investigation
can be extended to more realistic models, such as models with finite
chemical potential, isospin density, or the Sakai-Sugimoto
model~\cite{meson specfunction}. High spin meson can be described as
a rotating string in the gravity
background~\cite{Peeters:2006,Kruczenski:2003,Antipin:2007}.
\setcounter{equation}{0}
\subsection{Baryon in strongly coupled gauge field}

  A baryon has a configuration composed of $N_c$ fundamental strings
with the same orientation, which begin at the heavy quarks on the
boundary and end on the wrapped brane on the junction vertex in the
AdS$_5$~\cite{Witten:1998}. In the baryon configuration of a very
recent work~\cite{Liu:2008bs}, a D5 brane fills S$^5$ which sits a
point in AdS$_5$. A simple action of the hanging string and vertex
brane was used and the moving velocity dependence of screening
length was calculated, which is similar to the meson case.

In this paper, we consider a strings-junction configuration of
baryon in the finite temperature supersymmetric Yang-Mills dual of
the AdS-Schwarzschild black hole background. In particular, we
investigate the screening length for high spin baryons composed of
rotating $N_c$ heavy quarks. To rotate the heavy quarks by a finite
force, we put a hard infrared cutoff in the bulk and give the quark
a finite mass. We find that $N_c$ marcoscopic strings are embedded
reasonably in the bulk gravity with angular velocity $\omega$,
similar to the meson case. By defining the screening length as the
critical separation of quarks, we compute the $\omega$ dependence of
the screening length and find that rotating quarks can dissociate
more easily because of the centrifugal force. In section 2, we
analyze the baryon configuration in the AdS-Schwarzschild black hole
background, and find the strings+brane representation of the high
spin baryon in the bulk gravity. In section 3, we find the
interesting trajectory of strings in bulk space with angular
velocity $\omega$ and define the screening length of baryon by using
the usual method in a hot plasma. Then we analyze the interaction
potential between the quarks and show the critical behavior at the
screening length. Finally we get a reasonable $\omega$ dependence of
baryon screening length. In section 4, we relate the angular
momentum $J$ of $N_c$ strings to the high spin of a baryon and
defined the energy $E$ and $J$ charge. We pick up some
configurations with different $\omega$ and analyze the relation
between $E^2$ and $J$.

\setcounter{equation}{0}
\section{Baryon configuration}

We analyze a baryon composed of $N_c$ heavy probe quarks in the SYM
plasma at nonzero temperature. While on the supergravity side, there
are $N_c$ fundamental strings with the same orientation, which begin
at the heavy quarks on the boundary and end on the junction vertex
in the AdS$_5$. To give the quarks finite $\omega$, we need to give
them a finite mass scale at first, which means that we should put an
infrared cutoff in the bulk. We denote the boundary radius
$r_{\Lambda}$, then a free quark mass is \be m_q={1 \ov
2\pi\alpha'}(r_{\Lambda}-r_0)\;. \ee where the $1 \ov 2\pi\alpha'$
is the string tension and $r_0$ is the black hole horizon. A free
quark can be introduced as a string lying in the radial direction
from the boundary to the horizon. Also, an infrared cutoff in the
bulk can be realized by introducing a single brane, but for
simplicity we just ignore the backreaction and the dynamics of the
boundary brane in our paper. The vertex is a D5 brane wrapped on the
$S^5$ which sits at the same point in the AdS$_5$. The strings also
sit the same point in the $S^5$.

The dual background AdS$_5$$\times$S$^5$ metric with a black hole
is:
 \be
 ds^2 =  - f(r)  dt^2 +  \frac{r^2}{ R^2} d\vec x^2 +  \frac{dr^2}{f(r)} + R^2 d \Omega_5^2 \;,
 \ee
where
  \be
f(r)=\frac{r^2}{R^2}\left(1-\frac{r_0^4}{r^4}\right).
 \ee
and $R$ is the curvature radius of the AdS metric, $r$ is the
coordinate of the 5th dimension of AdS$_5$ and $r_0$ is the position
of the black hole horizon, $d \Omega_5^2$ is the metric for a unit
$S_5$. The temperature of the gauge theory is given by the Hawking
temperature of the black hole, $T=r_0/(\pi R^2)$. The gauge theory
parameters $N_c$ and $\lambda$ are given by \be \sqrt{\lambda}={R^2
\ov \alpha'}\;, \qquad
 {\lambda \ov N_c}= g^2_{\rm YM}= 4\pi g_s \;,
\ee where $g_s$ is the string coupling constant and ${1
\ov2\pi\alpha'}$ is the string tension. Infinite $N_c$ and large
$\lambda$ correspond to large string tension and weak string
coupling and thus justify the classical gravity treatment. We define
the dimensionless quark mass \be M_q={m_q \ov
T}={(r_{\Lambda}-r_0)R^2 \ov
2\alpha'r_0}=\sqrt{\lambda}{r_{\Lambda}-r_0 \ov 2r_0} \;. \ee When
quarks move in a hot plasma, the plasma is like wind if we stand
still in the rest frame of quarks. The boosted metric can be used to
calculate the velocity dependence of the screening
length~\cite{Liu:2008bs}. In our paper, we consider quarks rotating
in the $x_1-x_2$ plane, the background metric can be written as
\begin{equation}\label{boostmetric}
ds^2=-f(r)dt^2 +{r^2\ov R^2}dx_3^2 +\frac{r^2}{R^2}\left(
d\rho^2+\rho ^2d\theta^2\right)+\frac{1}{f(r)}dr^2+R^2 d\Omega^2_5
\end{equation}
where
%
%
$\rho$ and $\theta$ are
the coordinates in $x_1-x_2$ plane.

Now, we consider a rotating strings configuration corresponding a
baryon with high spin. It is composed of $N_c$ rotating quarks
arranged in a circle on the boundary. For simplicity, we define the
same angular velocity of $N_c$ strings as constant $\omega$. The
axis lies along radial direction of the AdS and passes through the
central of the boundary circle. Then, we parametrize one of $N_c$
strings as
 \be \label{parameters}\tau = t \;, \qquad \sigma=r\;, \qquad
\theta =\omega t\;, \qquad \rho=\rho(r)\;.
 \ee
Due to symmetry, $N_c$ strings have the same embedding function
$\rho(r)$. The single string Nambu-Goto action is
\begin{equation}
S_{string}=\frac {1}{2 \pi \alpha' }\int d\sigma d\tau
\sqrt{-\det[h_{ab}]} \;,
\end{equation}
where $h_{ab}=g_{\mu\nu}{\partial x^\mu \partial x^\nu \ov
\partial \sigma^a \partial \sigma^b}$.
We consider zero moving velocity but a finite rotating speed. The
$v_{x_3}$ dependence of baryon properties can be read
from~\cite{Liu:2008bs}. The induced metric on the world sheet
$h_{ab}$ is time independent, and the action can be written as
\begin{equation}\label{NGaction}
S_{string}=\frac {\TT}{2\pi \alpha'}\int_{r_e}^{r_{\Lambda}}dr
\sqrt{-\left(\frac{r^2}{R^2}\rho^2\omega^2-f(r)\right)
\left(\frac{1}{f(r)}+\frac{r^2}{R^2}\rho'^2(r)\right)}\;,
\end{equation}
where $\TT$ is the total time. The action for the D5 brane can be
written as~\cite{Liu:2008bs}
\be
 S_{D5} = \frac {\VV (r_e) \TT V_5 }{(2 \pi)^5 \alpha'^3 }\;,
\ee where $V_5$ is the volume of the compact brane and
$\VV(r_e)=\sqrt{-g_{00}}$ is the potential for the brane located at
$r=r_e$. Since the D5 brane sits at a point in the AdS space, we
ignore the $\omega$ dependence of the $S_{D5}$.

The total action of the system is then
 \be
  S_{total} =
  \sum_{i=1}^{N_c}  S_{string}^{(i)} + S_{D5} \;,
 \ee
where we can consider only one typical string to get $N_c$ strings'
action for symmetry. We find static baryon configuration through
extremizing $S_{total}$, first with respect to the $\rho(r)$(we
ignore the $x_3$ and $\theta$ for symmetry) that describes the
embedding of the $N_c$ strings, and then with respect to $\rho(r_e)$
and $r_e$, the location of the vertex brane.

Generally, extremizing $S_{total}$ with respect to $\rho(r_e)$ and
$r_e$ yields the $x_1-x_2$ plane and $r$ directions force balance
condition(FBC). Since the system is axial symmetric, the position of
baryon vertex $\rho(r_e)$ is fixed and treated as the origin in the
$x_1-x_2$ plane. But bulk coordinate $r_e$ is nonzero and believed
to be dependent on $\omega$.

As above, we get the radial FBC by extremizing $S_{total}$ with
respect to $r_e$

 \be\label{FBC}
  \sum_{i=1}^{N_c} H^{(i)} \biggr|_{r_e} = \Sigma \;,
 \ee
 where
 \be\label{H}
 H^{(i)} \equiv
 \LL^{(i)} -  \rho'^{(i)} {\p \LL^{(a)} \ov \p \rho'^{(i)}} \;,
 \ee

\be\label{sigma}
 \Sigma \equiv {2 \pi \apr \ov \TT} {\p S_{\rm D5} \ov \p
   r_e}  = {V_5 \ov (2 \pi)^4 \apr^2} {\p \VV(r_e) \ov \p r_e} \;,
\ee

In the following section, we will find the shape of the hanging
string in the rotating case and the relation between quark
separation $l_q$ and $r_e$ change a lot when we increase $\omega$
from zero. The $\omega$ dependence of embedding function $\rho(r)$
and the spin dependence of baryon screening length will have
apparent physical explanation.

\setcounter{equation}{0}
\section{$\omega$ dependence of the screening length}

\begin{figure}[t]
\begin{minipage}[t]{0.5\linewidth}
\centering
\includegraphics*[width=0.85\columnwidth]{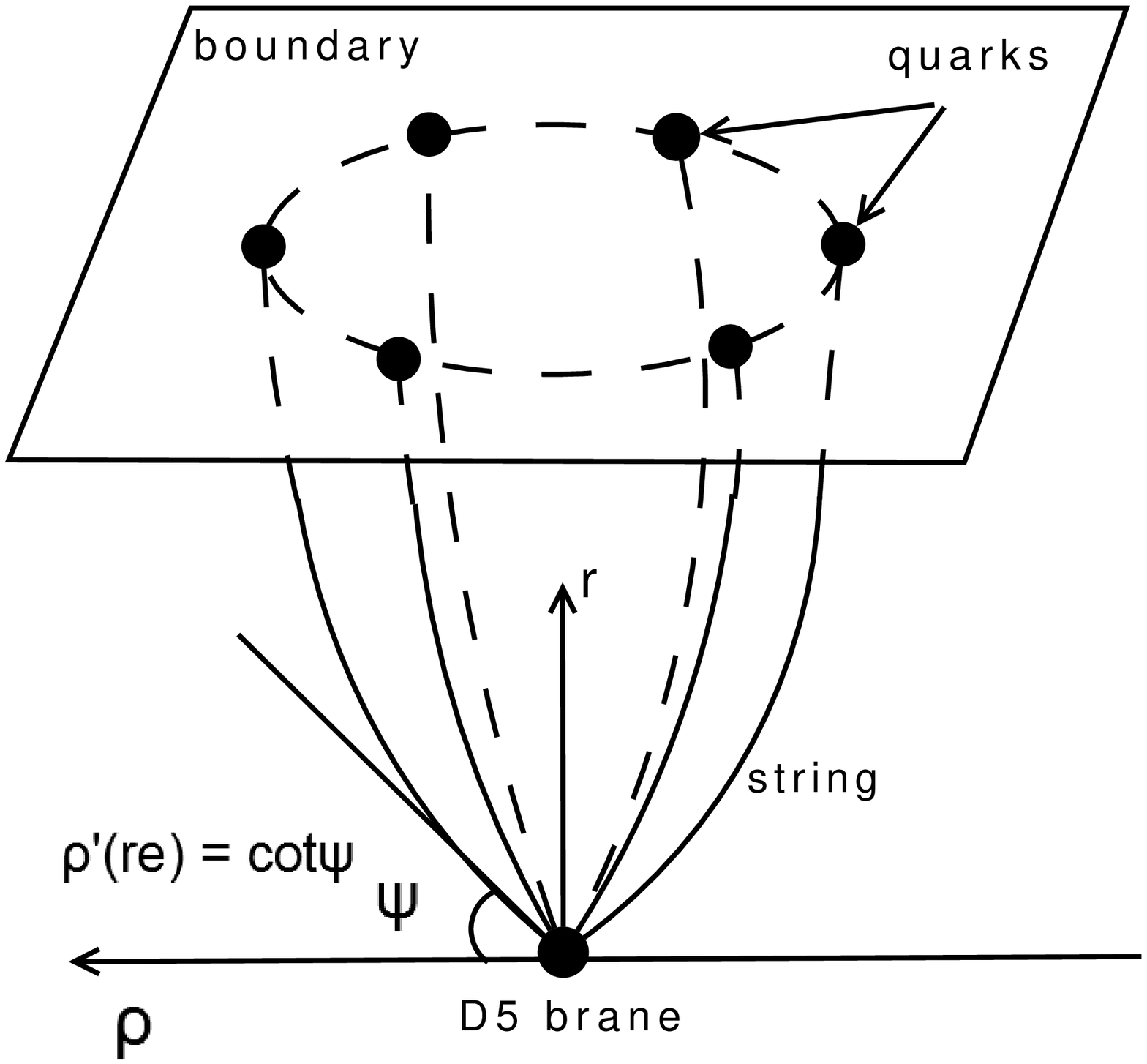}
\caption{\small Strings+brane representation of a baryon.
}\label{fig:01}
\end{minipage}
\begin{minipage}[t]{0.5\linewidth}
\centering
\includegraphics*[width=0.7\columnwidth]{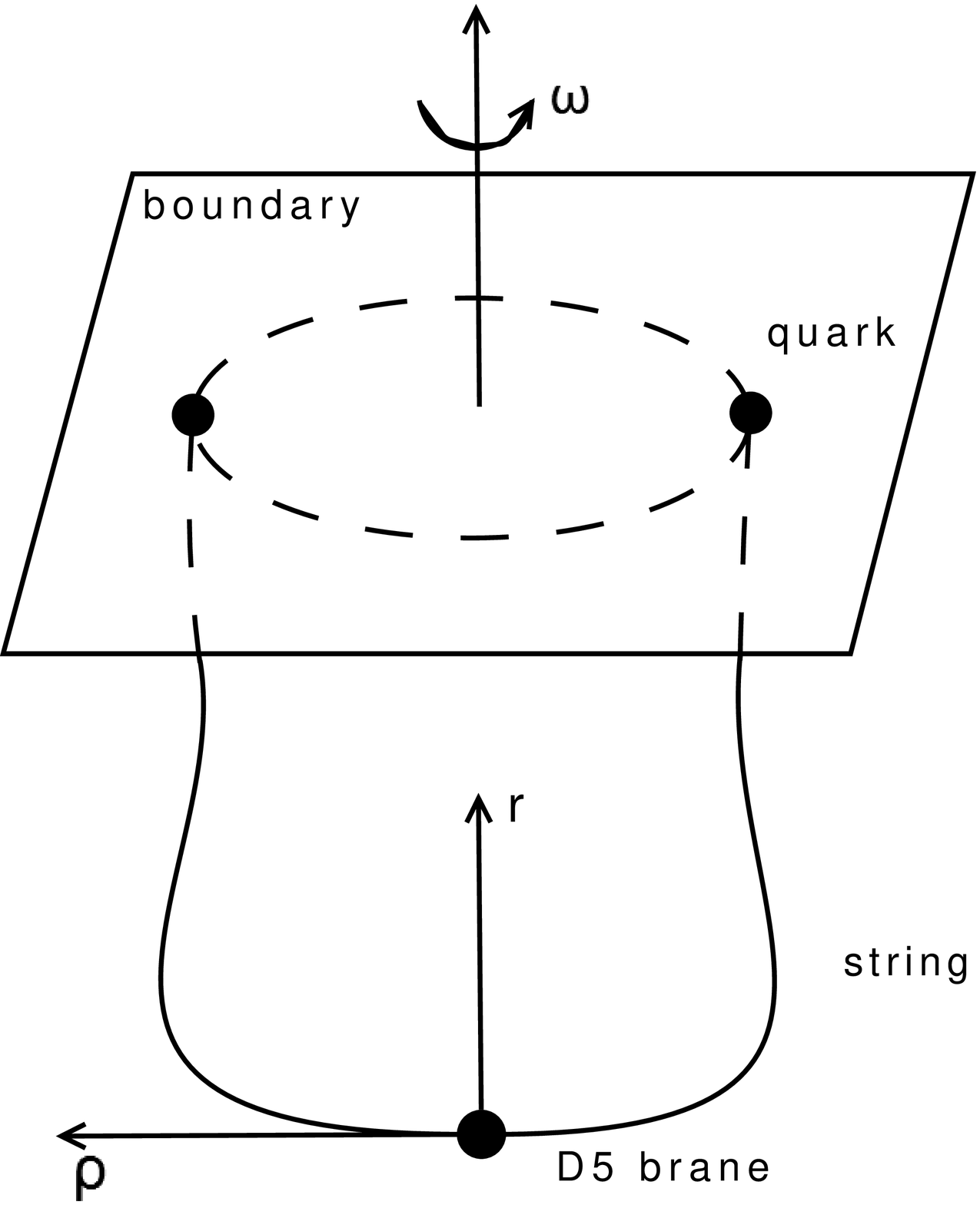}
\caption{\small Strings+brane representation of a high spin baryon,
the bottoms of the strings corresponds to the cusp of the embedding
function in Figure \ref{fig:rho-r}.}\label{fig:02}
\end{minipage}
\end{figure}

To find the static baryon configuration, we extremize $S_{total}$
with respect to $\rho(r)$ describing trajectories of the $N_c$
strings in AdS. The world sheet Lagrange density with parameters in
(\ref{parameters}) is

\be \LL=\sqrt{-\left(\frac{r^2}{R^2}\rho^2\omega^2-f(r)\right)
\left(\frac{1}{f(r)}+\frac{r^2}{R^2}\rho'^2(r)\right)} \;. \ee The
equation of motion for $\rho(r)$:
\begin{equation}\label{EOM}
\left({\p\ov\p\rho(r)}-{\p \ov \p r}{\p\ov\p\rho'(r)}\right)\LL=0
\;.
\end{equation}
To solve the above equation of motion, we need two boundary
conditions, the values of $\rho(r_e)$ and $\rho'(r_e)$. Due to the
axial symmetry, $\rho(r_e)$ must vanish. We have $\rho(r_e)=0$ even
if we rotate the hanging strings. However, $\rho'(r_e)\neq0$ and it
need to be determined by other conditions. When we rotate the
massive strings, the centrifugal force will change the angle $\psi$
shown in figure \ref{fig:01}, where $\cot \psi=\rho'(r_e)$.


In fact we determine $r_e$ by the radial FBC at first. From
(\ref{FBC})(\ref{H})(\ref{sigma}), the FBC can be written as
 \be
 \sum_{i=1}^{N_c}(\LL^{(i)} -  \rho'^{(i)} {\p \LL^{(i)} \ov \p \rho'^{(i)}})={V_5 \ov (2 \pi)^4 \apr^2} {\p \VV(r_e) \ov \p
 r_e} \;.
 \ee
Due to the axial symmetry, we can write
 \be
 \LL -  \rho' {\p \LL \ov \p \rho'}={1 \ov N_c}{V_5 \ov (2 \pi)^4 \apr^2} {\p \VV(r_e) \ov \p
 r_e} \;.
 \ee
The above equation does not include any unknown function since we
choose $r=r_e$. This force balance condition gives
\be\label{FBC1}
 \rho'(r_e)=\frac{R}{r_e f(r_e)^{1/2}} \sqrt{\frac{R^2(r_e^4-r_0^4)r_e^4}{(r_e^4+r_0^4)^2
 A^2}-1}\;,
 \ee
 where
 \be
  A={1 \ov N_c}{V_5 \ov (2 \pi)^4\apr^2}
 \ee
 This equation gives rise to the relation between
$\rho'(r_e)$ and $r_e$. Because $\rho(r)$ depends on $\rho'(r_e)$,
we can solve the two equations (\ref{EOM}) (\ref{FBC1}) numerically
together.

\begin{figure}
\begin{minipage}[t]{0.5\linewidth}
\centering
  \includegraphics*[width=1.1\columnwidth]{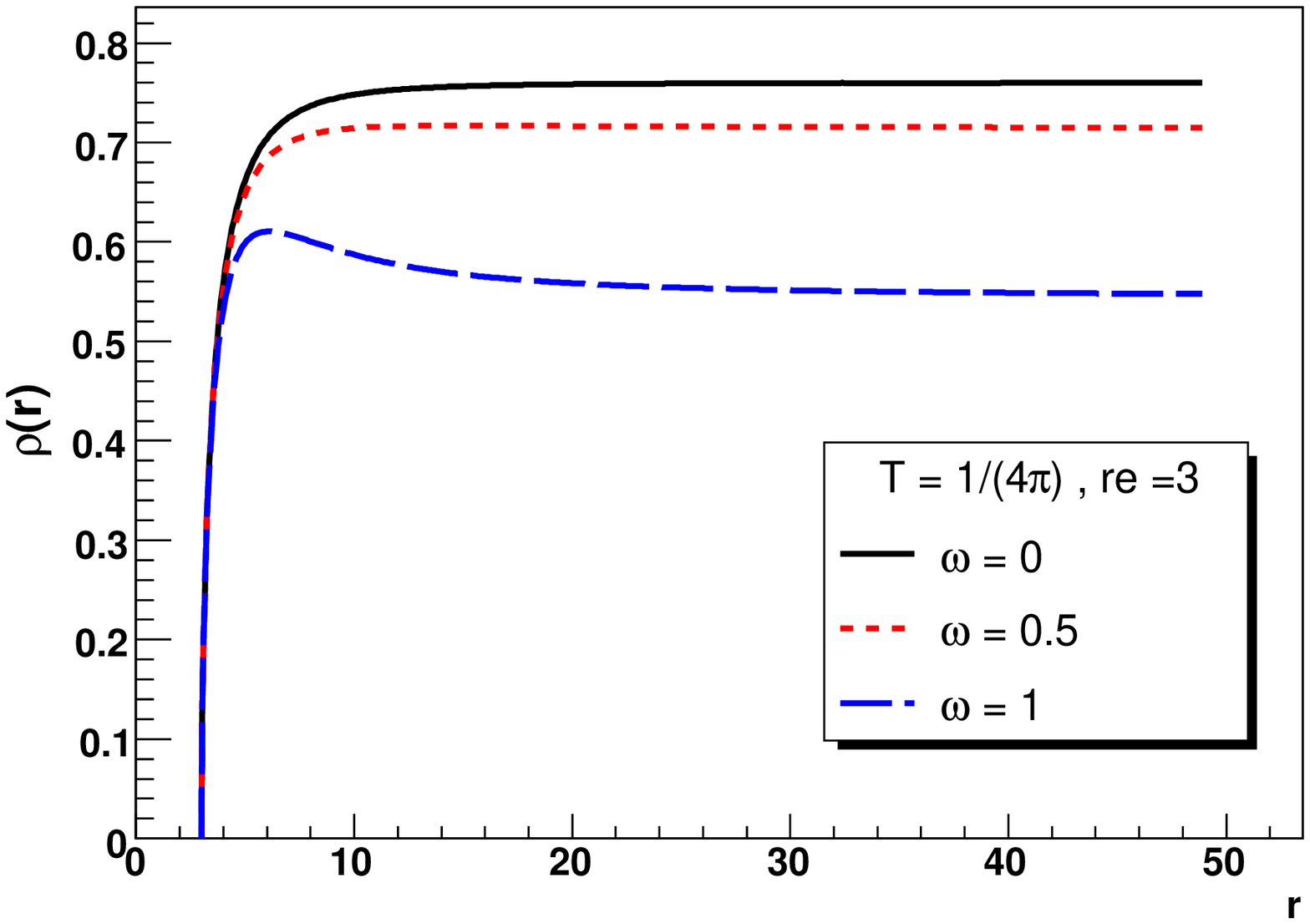}
  \caption{\small The embedding function $\rho(r)$ at different values of
  $\omega$ with fixed $r_e$.}\label{fig:rho-r}
\end{minipage}
\begin{minipage}[t]{0.5\linewidth}
\centering
\includegraphics*[width=1.1\columnwidth]{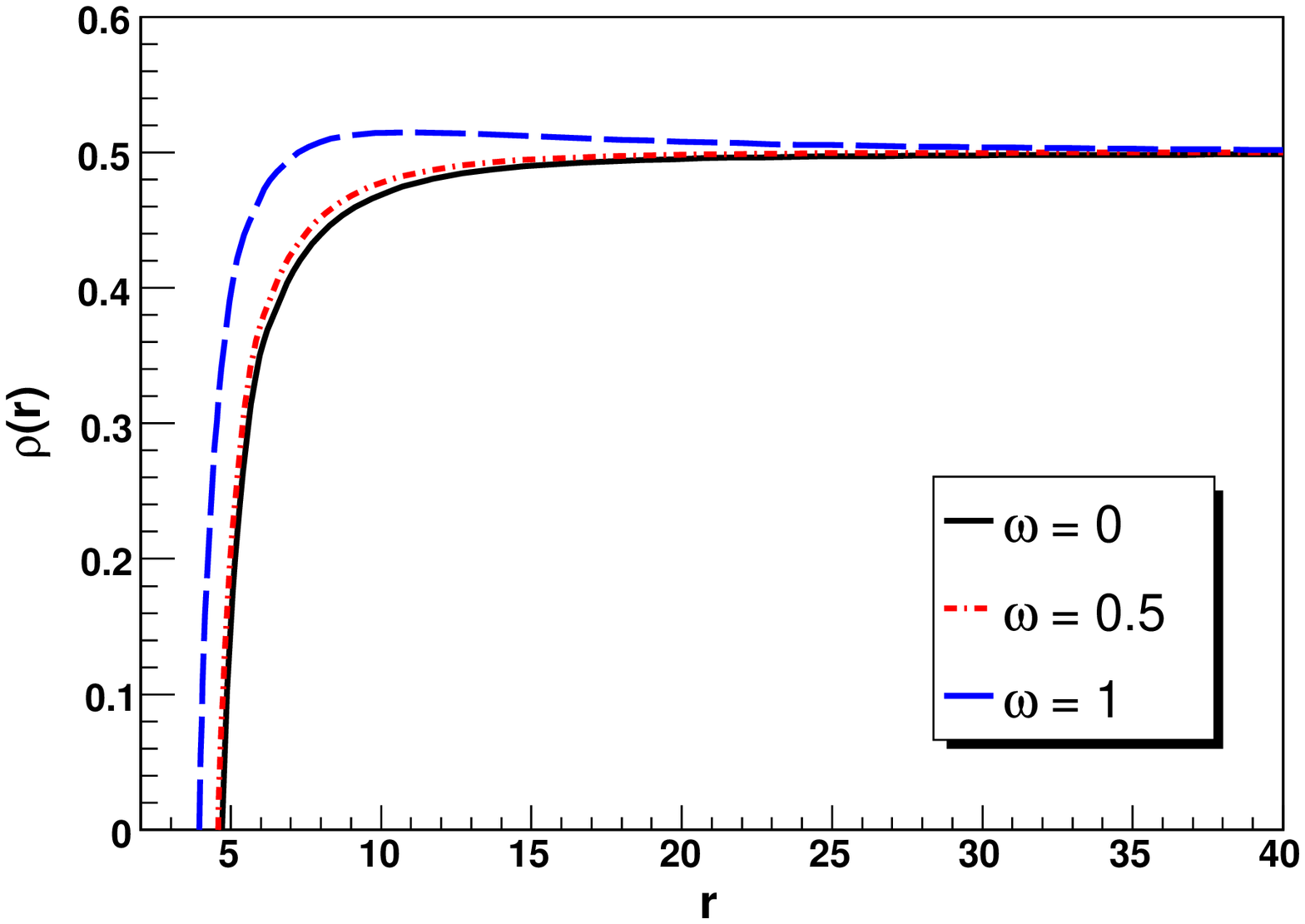}
\caption{\small The embedding function $\rho(r)$ at different values
of $\omega$ with fixed $l_q$,
$r_{\Lambda}=100$.}\label{fig:rho_r_lq}
\end{minipage}
\end{figure}




Solving equation (\ref{FBC1}), we have one free input value of
$r_e$. It also means that we can input a length of radial trajectory
of $N_c$ strings $r_{\Lambda}-r_e$. When $r_e$ is given, the quark
separation on the cutoff brane $l_q$ can be calculated by the
solution $\rho(r)$ to

\be
 l_q=2\int_{r_e}^{r_{\Lambda}}\rho'(r) dr=2\rho(r_{\Lambda}) \;.
\ee


When $\omega$ increases from 0 to a finite value, the embedding
function $\rho(r)$ can be solved, shown in Figure \ref{fig:rho-r}.
In this figure, we choose the embedding functions of different
baryons with the same $r_e$. Here, by ``different'', we mean that
the initial separations between quarks and their energy are
different even when we have not rotated them. It also means that the
interaction potential between quarks are different while $\omega=0$
for the curves in Figure \ref{fig:rho-r}. For baryon of same spin a
larger interaction potential also corresponds to a larger boundary
quark separation, which will be shown in Figure \ref{fig:EI-lq} in
the next section. Figure \ref{fig:rho-r} teaches us that the strings
warp near the horizon but become straight near the boundary, it's
reasonable because the compact D5 brane is like a box which
transfers the interaction between the strings. The corresponding
configuration of the baryon is shown in Figure \ref{fig:02}. The tip
of the string becomes lower when we increase $\omega$ , to guarantee
that the speed of the tip is smaller than the speed of light and to
make the centralfugal force not too large. Since the bottom part of
a hanging string is heavy, at the same centrifugal acceleration, it
sticks out in the $\rho$ direction apparently.

\begin{figure}
\begin{minipage}[t]{0.5\linewidth}
\centering
\includegraphics*[width=1.1\columnwidth]{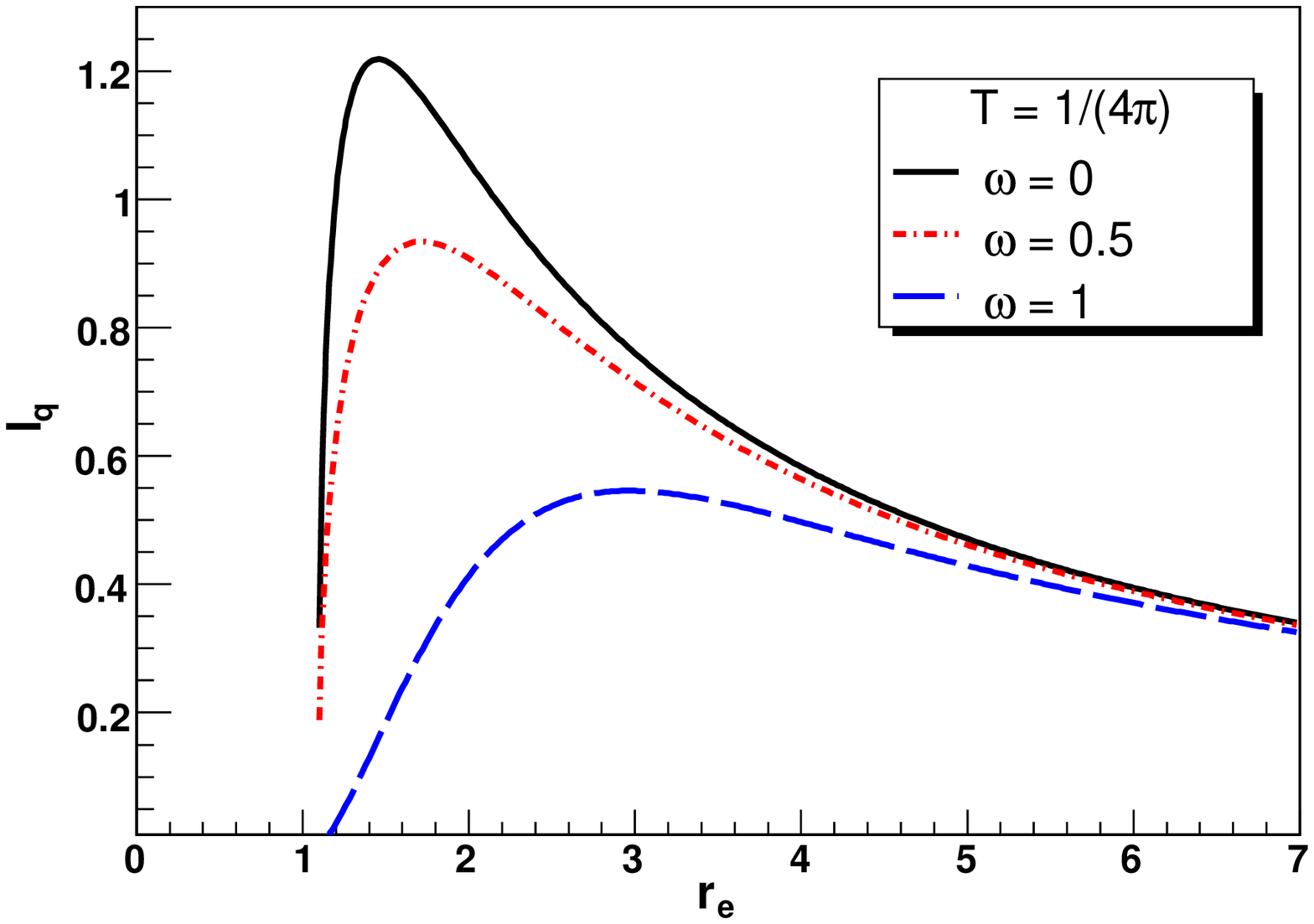}
\caption{\small $r_e$ dependence of $l_q$ at different $\omega$.
Each curve has two branches and we choose the right part as the
physical baryons, $r_{\Lambda}=100$.}\label{fig:lq-re2}
\end{minipage}
\begin{minipage}[t]{0.5\linewidth}
\centering
\includegraphics*[width=1.1\columnwidth]{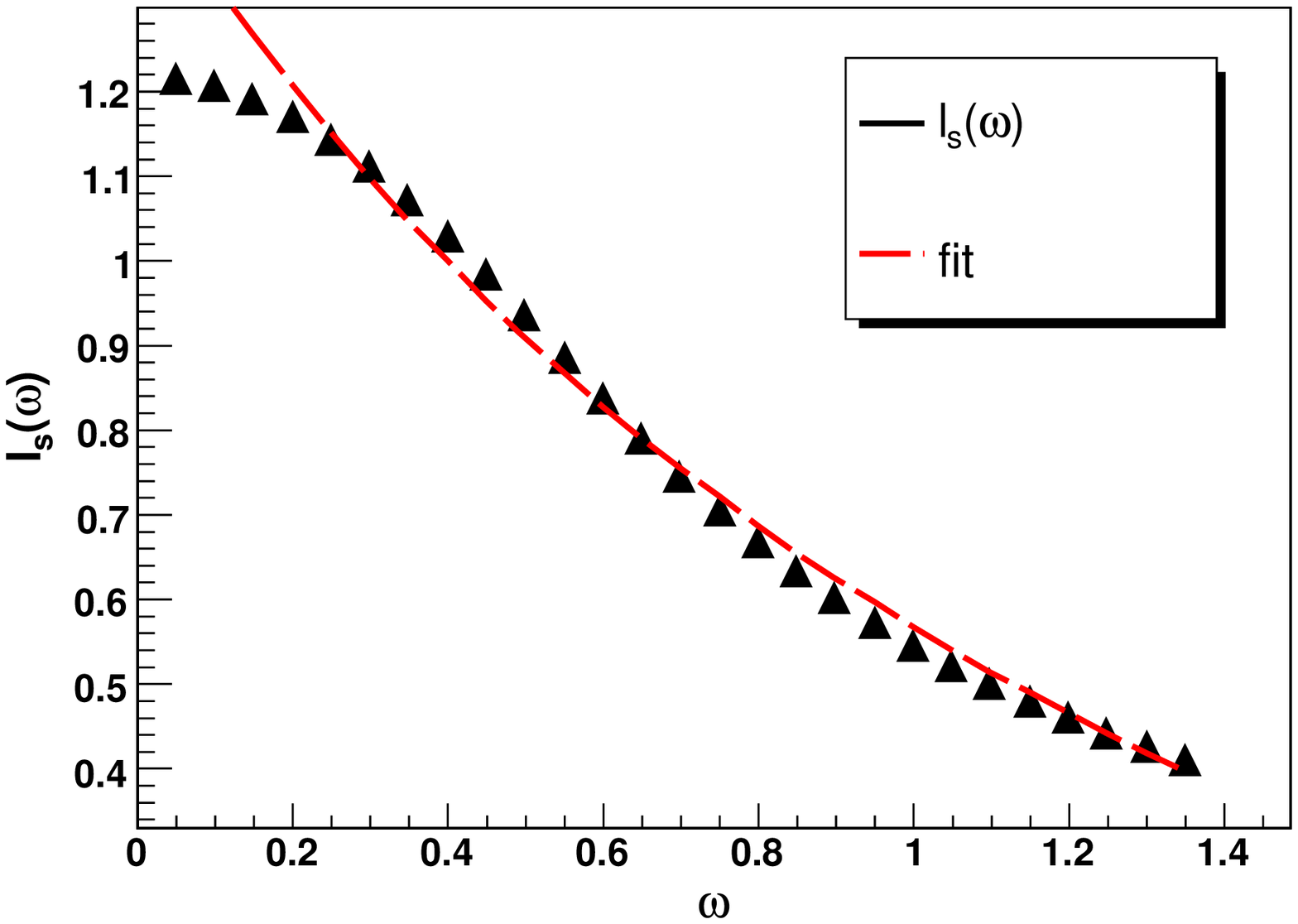}
\caption{\small $\omega$ dependence of the screening length. The fit
curve is the modified inverse proportion function,
$r_{\Lambda}=100$.}\label{fig:ls}
\end{minipage}
\end{figure}



If we fix the separation of boundary quarks and stand in the static
frame of baryon, the rotating medium will try to separate them and
make connecting strings in the bulk longer. We see these solutions
in Figure \ref{fig:rho_r_lq}. We are interested in the behavior of
$l_q$ as $r_e$ changes. For different values of $\omega$, we plot
$r_e$ dependence of $l_q$ in Figure \ref{fig:lq-re2}. In this
figure, there is an important phenomenon that $l_q$ becomes large at
first and then turns to be smaller, as we increase $r_e$. Actually,
for a given $l_q$, we get two possible values of $r_e$ from curves
in this figure. We consider that the right part of the each curve
stands for real baryon states and the left part is unstable and will
be thrown away. It is reasonable because in the right part, $r_e$
becomes smaller when $l_q$ becomes larger, which means that a high
energy baryon can probe deeper position in the bulk than a low
energy baryon if they have same spin. We will see that the energy of
a stable baryon with fixed spin must be larger if the average
separation between the quarks is larger in the next section. In
another view, for baryons of same spin, the only way to have high
energy is to keep the quarks with large separation stable. However,
we can introduce high spin to revise this relation between $l_q$ and
$r_e$ as shown in this figure. High spin helps to weaken the
increasing of $l_q$ as $r_e$ becomes smaller. In our classical
picture, rotating strings can contribute more energy than unrotating
strings.

\begin{figure}[t]
\begin{center}
\includegraphics[width=.9\columnwidth]{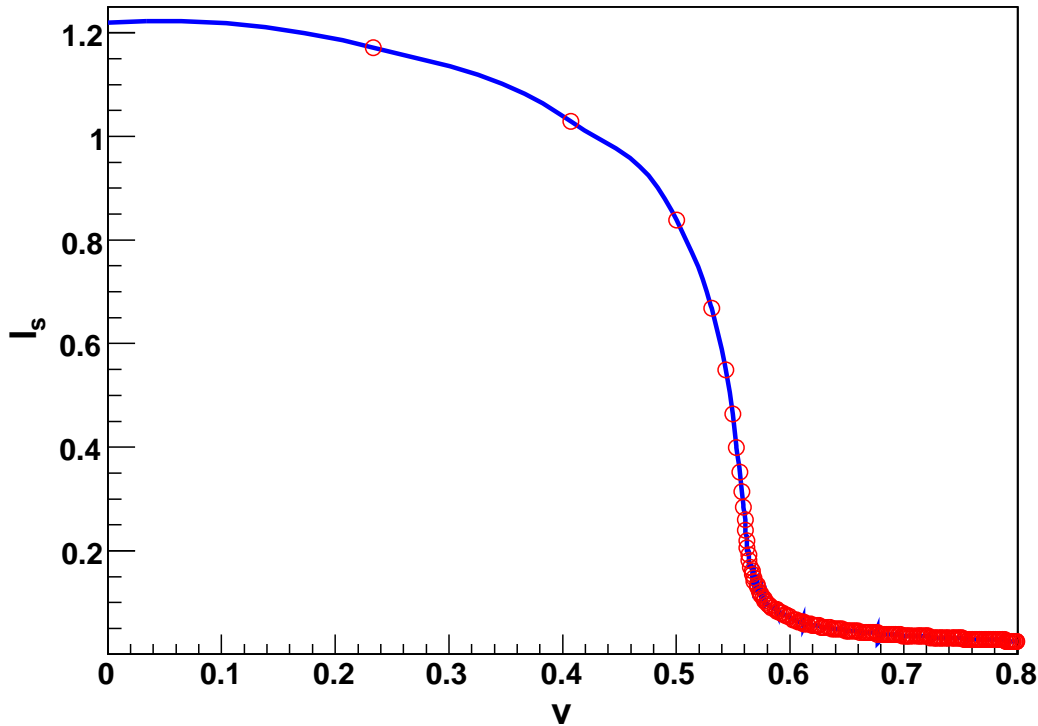}
\end{center}
\caption{Linear velocity dependence of screening length}\label{ls-v}
\end{figure}

\begin{figure}
\begin{minipage}[t]{0.48\linewidth}
\centering
\includegraphics*[width=1.0\columnwidth]{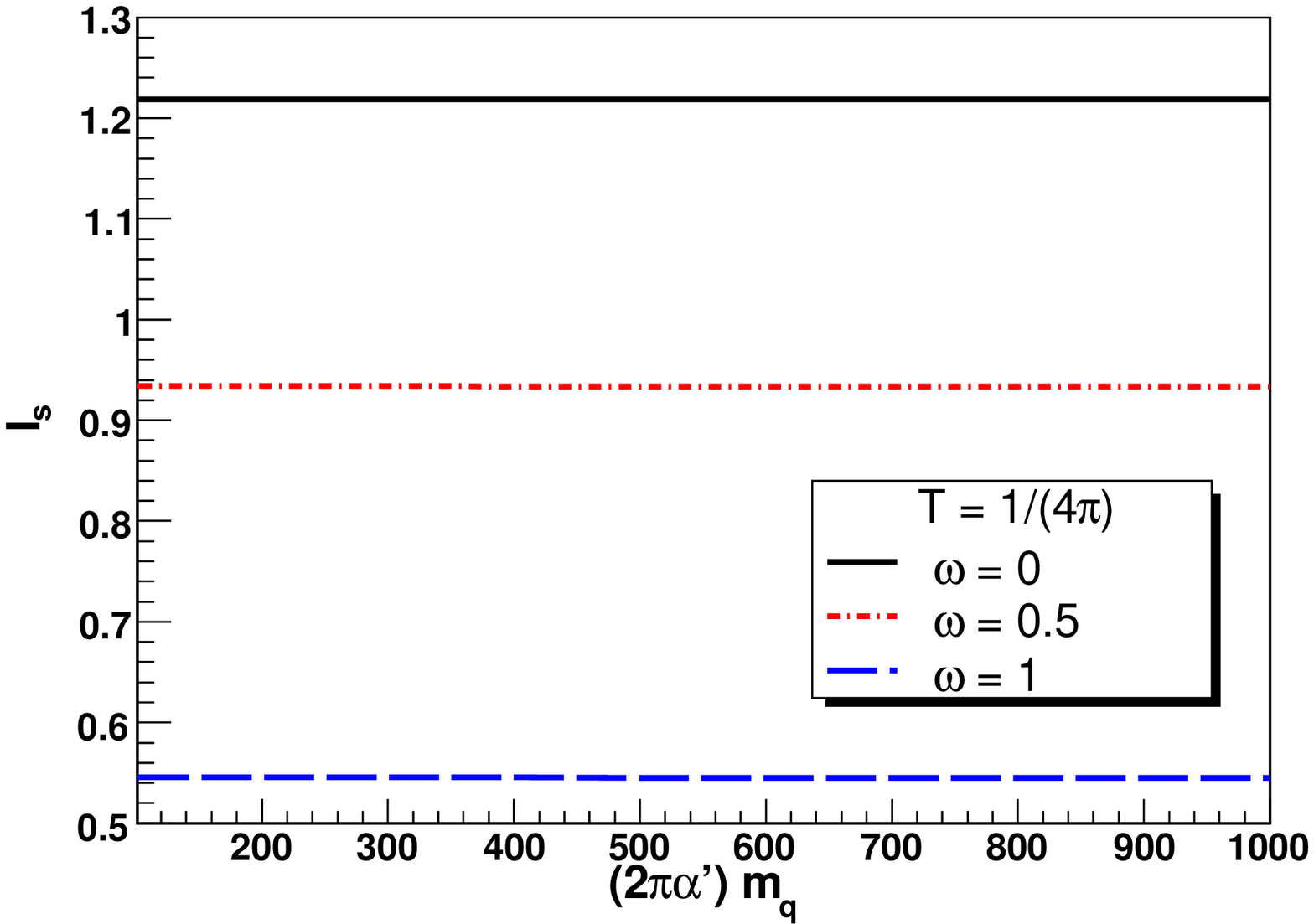}
  \caption{\small $r_{\Lambda}$ dependence of screening length at different values of
  $\omega$\;.
    }\label{fig:ls-cut}
\end{minipage}
\begin{minipage}[t]{0.48\linewidth}
\centering
\includegraphics*[width=1.0\columnwidth]{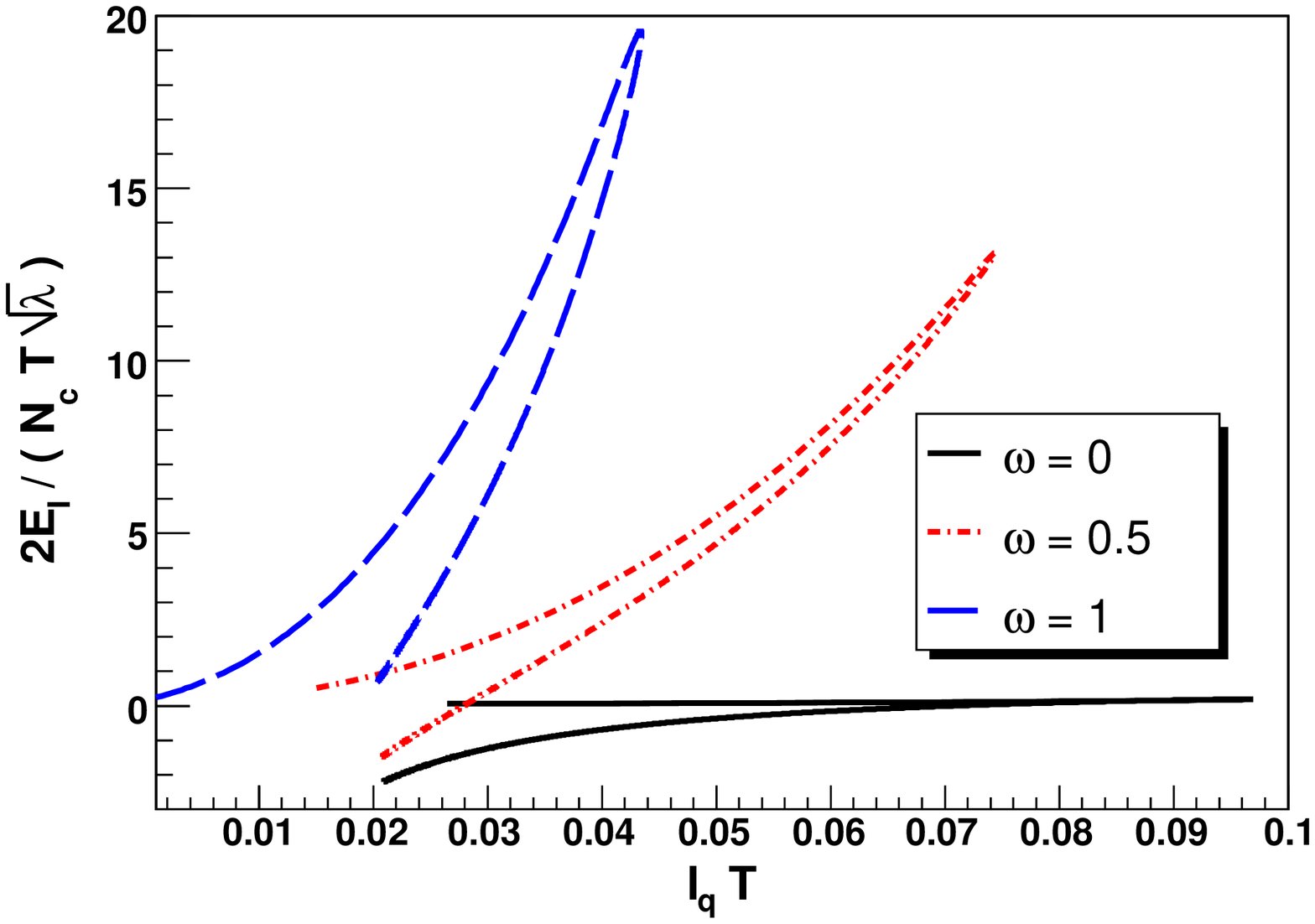}
  \caption{\small The interaction potential of baryon configuration with a given
  $l_q$ for several values of the angular velocity $\omega$, $r_{\Lambda}=100$.  Two
  branches meet at $l_q=l_s$\;.
    }\label{fig:EI-lq}
\end{minipage}
\end{figure}

We define the baryon screening length as the critical quark
separation in Figure \ref{fig:lq-re2}. When $\omega$ increases from
zero, the screening length seems to decrease as $l_s\propto{1\ov
\omega}$ in Figure \ref{fig:ls}. When we fit the curve using

\be
 l_s={a \ov b \omega+c}-d,
 \ee
we find the best result is \be l_s=\frac{0.51}{0.14
\omega+0.22}-0.85.
 \ee\label{fit_ls}
We also finished the numerical computation of $v$ dependence of $l_s$ in Figure \ref{ls-v}, where $v$ is the linear
velocity of quarks, where $v=l_q\omega/2$. When $\omega$ is very large, the linear velocity is indeed close to 1. At
the same time the quark separation is very small. Our calculation shows that $l_s$ decrease very slowly as the velocity
becomes close to 1. \footnote{ In our paper, $l_s$ is a parameter which shows how big the baryon can become to keep
alive in the plasma. It's different from the ``maximum spin size'' in \cite{Peeters:2006}.}
However, we can not find a suitable function like (\ref{ls_v}) to fit the curve. The curve is very different from the
usual curve which describes dragging velocity dependence of $l_s$ in (\ref{ls_v}). The main reason we think is that the
dragging effect is very different from the rotating effect. Actually, different parts of hanging string have different
linear velocities, which is very different from the dragging case. We should note that if we add a drag force in the
$x_3$ direction which is orthogonal to the rotating plane, we may find that $v_{x_3}$ dependence of $l_s$ is similar to
(\ref{ls_v})\cite{Peeters:2006}.

In hot plasma, the screening
length depends on the temperature directly and it is different for
various hadrons. The smaller screening length of a baryon with
higher spin makes itself dissociate more easily at a certain
temperature in the plasma. The real plasma always becomes cold very
quickly after formation, the baryons with high spin appear later
than the low spin baryons.


For simplicity, we ignore the backreaction of the cutoff brane. For
different values of $m_q$, the screening lengths are shown in Figure
\ref{fig:ls-cut}. In this figure, we find that the screening length
almost keeps the same value while we choose different cutoffs. It
shows that the screening length is a property of the medium, which
is a strongly coupled SYM in this case.
When we choose a very large cutoff in the bulk,
the bulk theory is more like the SYM. Due to conformal invariance of
SYM, scale dependence of screening length is negligible.

\setcounter{equation}{0}
\section{High spin baryon at nonzero temperature}
In section 3, we argue that a baryon with high spin can be described
by rotating semiclassical strings together with a massive brane.
Screening length is considered as the most important signal of quark
gluon plasma. Now, we will pay more attention to the properties of
the baryon itself. In this section, baryon energy and angular
momentum will be defined. More evidence of high spin will be given
by the relation of the energy and the angular momentum. We find two
branches of $J-E^2$ curves.

\begin{figure}[t]
\begin{minipage}[t]{0.48\linewidth}
\centering
\includegraphics*[width=0.9\columnwidth]{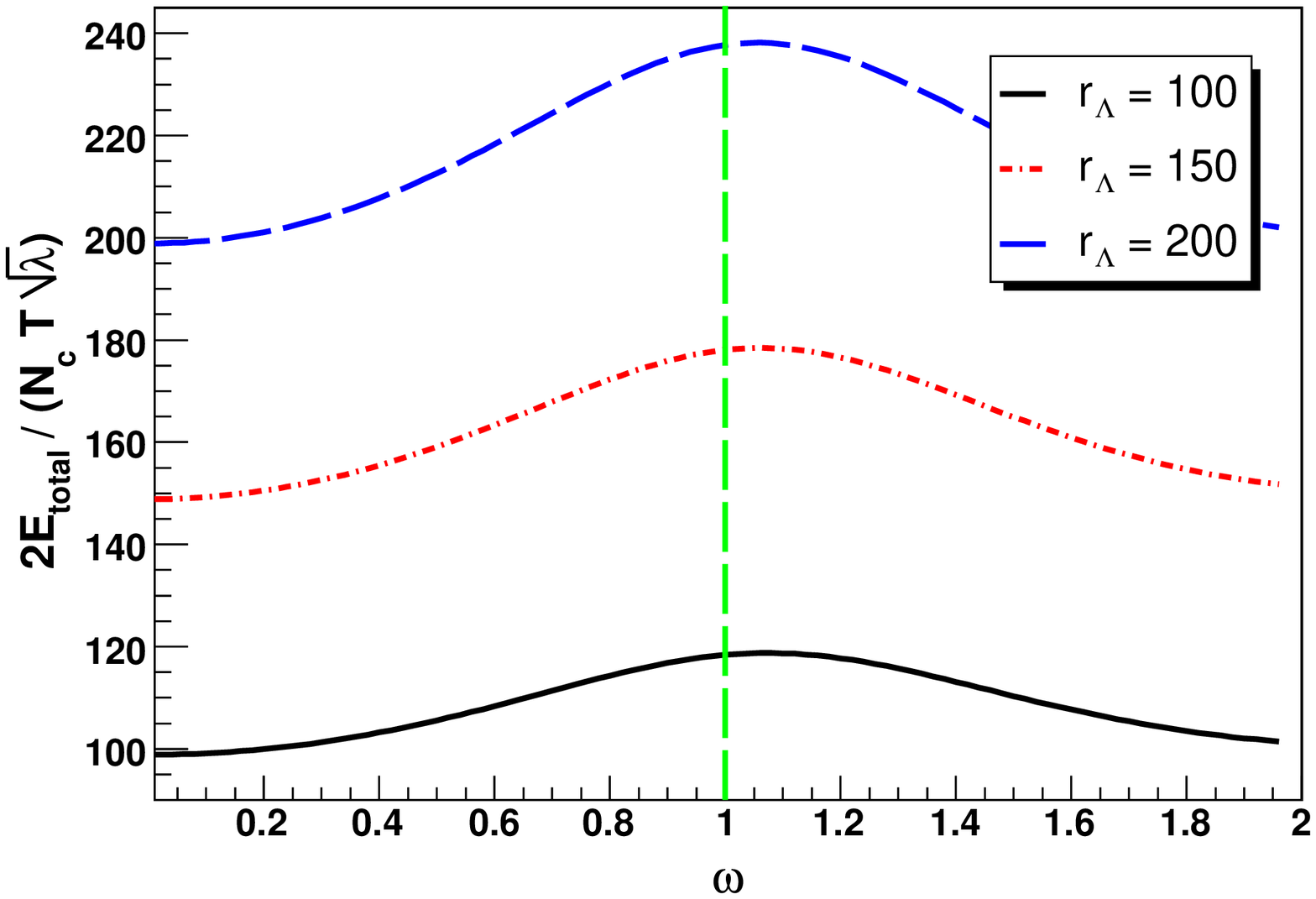}
  \caption{\small $\omega$ dependence of baryon energy with
  fixed
  $r_e$.
    }\label{fig:E-omega}
\end{minipage}
\begin{minipage}[t]{0.48\linewidth}
\centering
\includegraphics*[width=0.9\columnwidth]{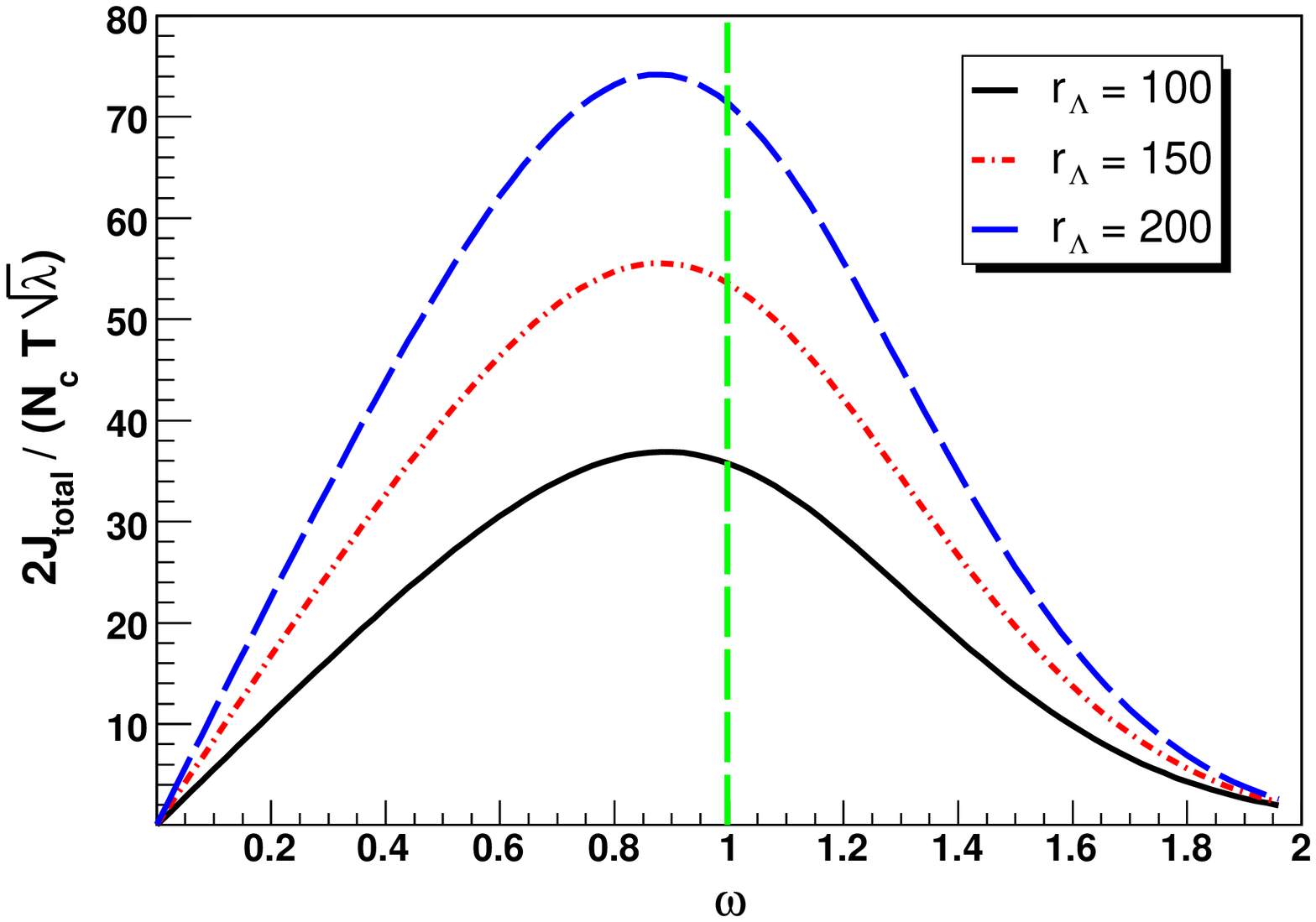}
  \caption{\small $\omega$ dependence of baryon $J$ charge with
  fixed
  $r_e$.
    }\label{fig:J-omega}
\end{minipage}
\begin{minipage}[t]{0.48\linewidth}
\centering
\includegraphics*[width=0.9\columnwidth]{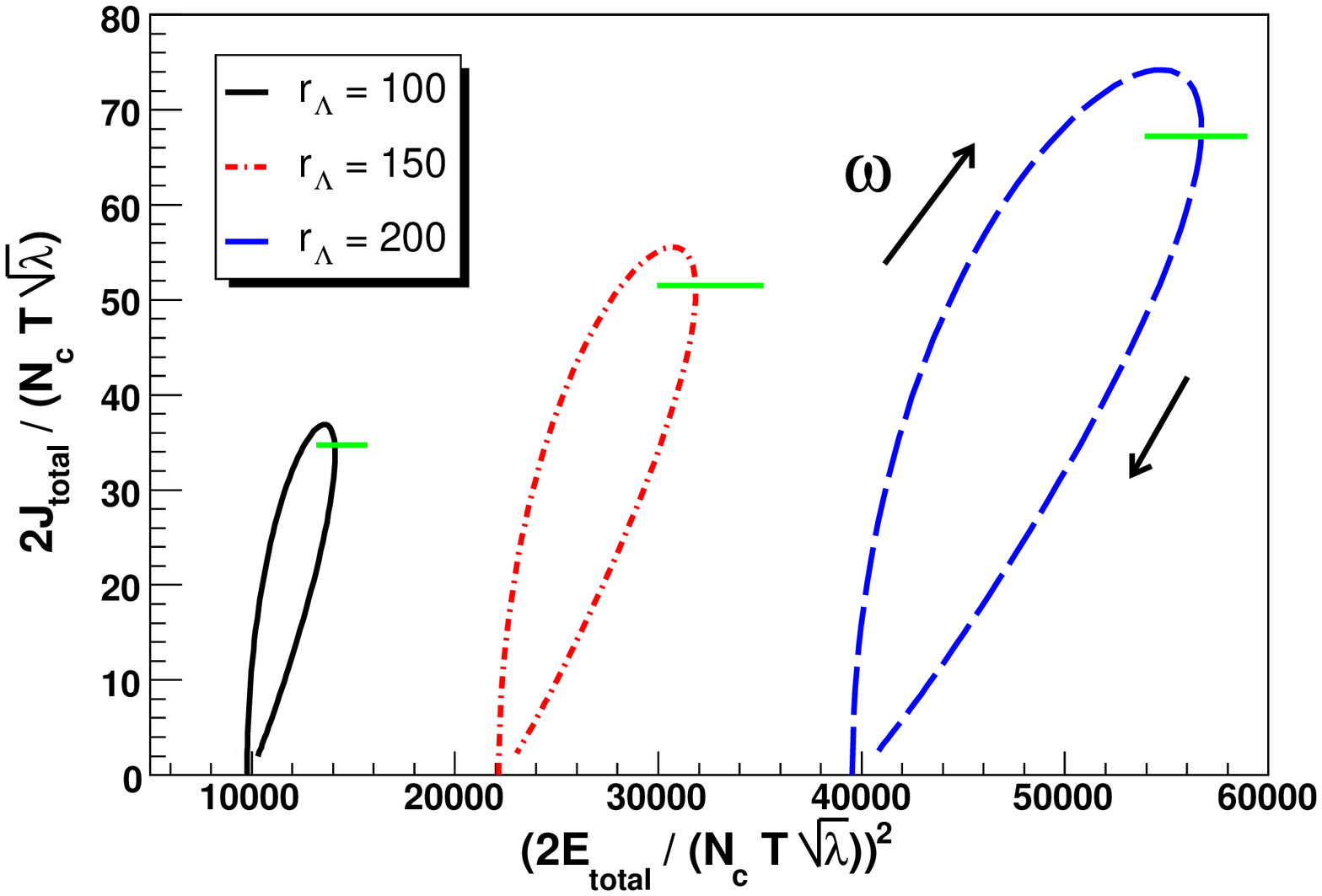}
  \caption{\small Two branches of $J-E^2$ behavior
  of baryon with
  fixed
   $r_e$.
    }\label{fig:regg}
\end{minipage}
\begin{minipage}[t]{0.48\linewidth}
\centering
\includegraphics*[width=0.9\columnwidth]{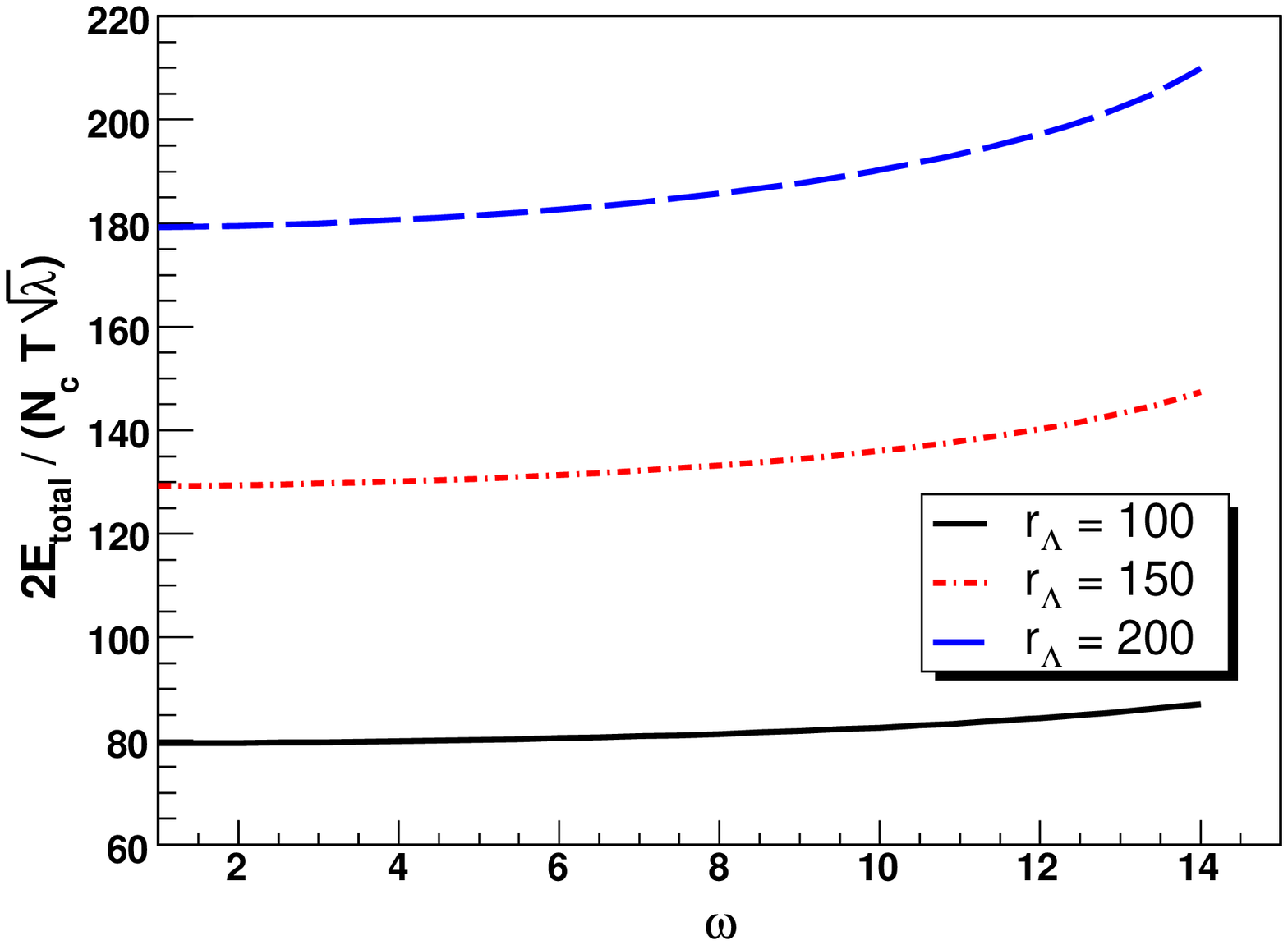}
  \caption{\small $\omega$ dependence of baryon energy with
  fixed
  $l_q$ on the boundary.
    }\label{fig:E_omega_lq}
\end{minipage}
\end{figure}
Now, we want to give the definition of baryon energy and angular
momentum. We write the string Lagrange from the Nambu-Goto action
(\ref{NGaction})
 \be L=\frac {1}{2\pi
\alpha'}\int_{r_e}^{r_{\Lambda}}dr
\sqrt{-\left(\frac{r^2}{R^2}\rho^2\omega^2-f(r)\right)
\left(\frac{1}{f(r)}+\frac{r^2}{R^2}\rho'^2(r)\right)}\;. \ee

The angular momentum and energy of the string (we choose positive
values) are
 \be\label{jcharge}
 J_{string}={\p L \ov \p\omega}=\frac {1}{2\pi
\alpha'}\int_{r_e}^{r_{\Lambda}}dr\frac{\left(\frac{1}{f(r)}+\frac{r^2}{R^2}\rho'^2(r)\right)(\frac{r^2}{R^2}\rho^2\omega)}
{\sqrt{-\left(\frac{r^2}{R^2}\rho^2\omega^2-f(r)\right)
\left(\frac{1}{f(r)}+\frac{r^2}{R^2}\rho'^2(r)\right)}}, \ee and
\be\label{stringenergy} E_{string}=\omega {\p L \ov
\p\omega}-L=\frac {1}{2\pi
\alpha'}\int_{r_e}^{r_\Lambda}dr\frac{\left(\frac{1}{f(r)}+\frac{r^2}{R^2}\rho'^2(r)\right)f(r)}
{\sqrt{-\left(\frac{r^2}{R^2}\rho^2\omega^2-f(r)\right)
\left(\frac{1}{f(r)}+\frac{r^2}{R^2}\rho'^2(r)\right)}}\;.
 \ee
The energy of D5 brane is given by
 \be\label{Ebrane}
 E_{brane}={{\VV(r_e)V_5} \ov (2\pi)^5\alpha'^3}\;.
\ee

\begin{figure}[t]
\begin{minipage}[t]{0.48\linewidth}
\centering
  \includegraphics*[width=1.0\columnwidth]{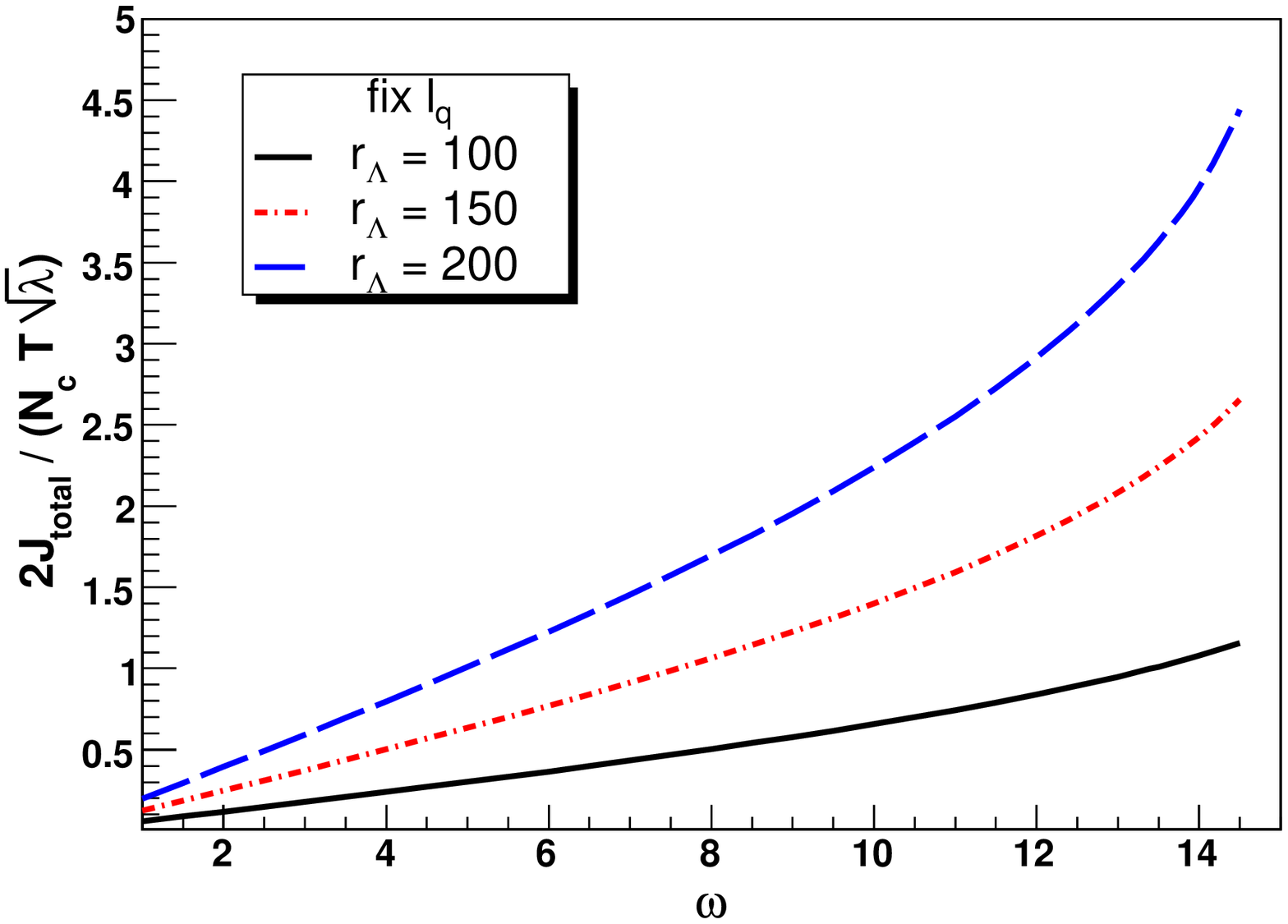}
  \caption{\small $\omega$ dependence of baryon $J$ charge with
  fixed
 $l_q$ on the boundary.
    }\label{fig:J_omega_lq}
\end{minipage}
\begin{minipage}[t]{0.48\linewidth}
\centering
  \includegraphics*[width=1.0\columnwidth]{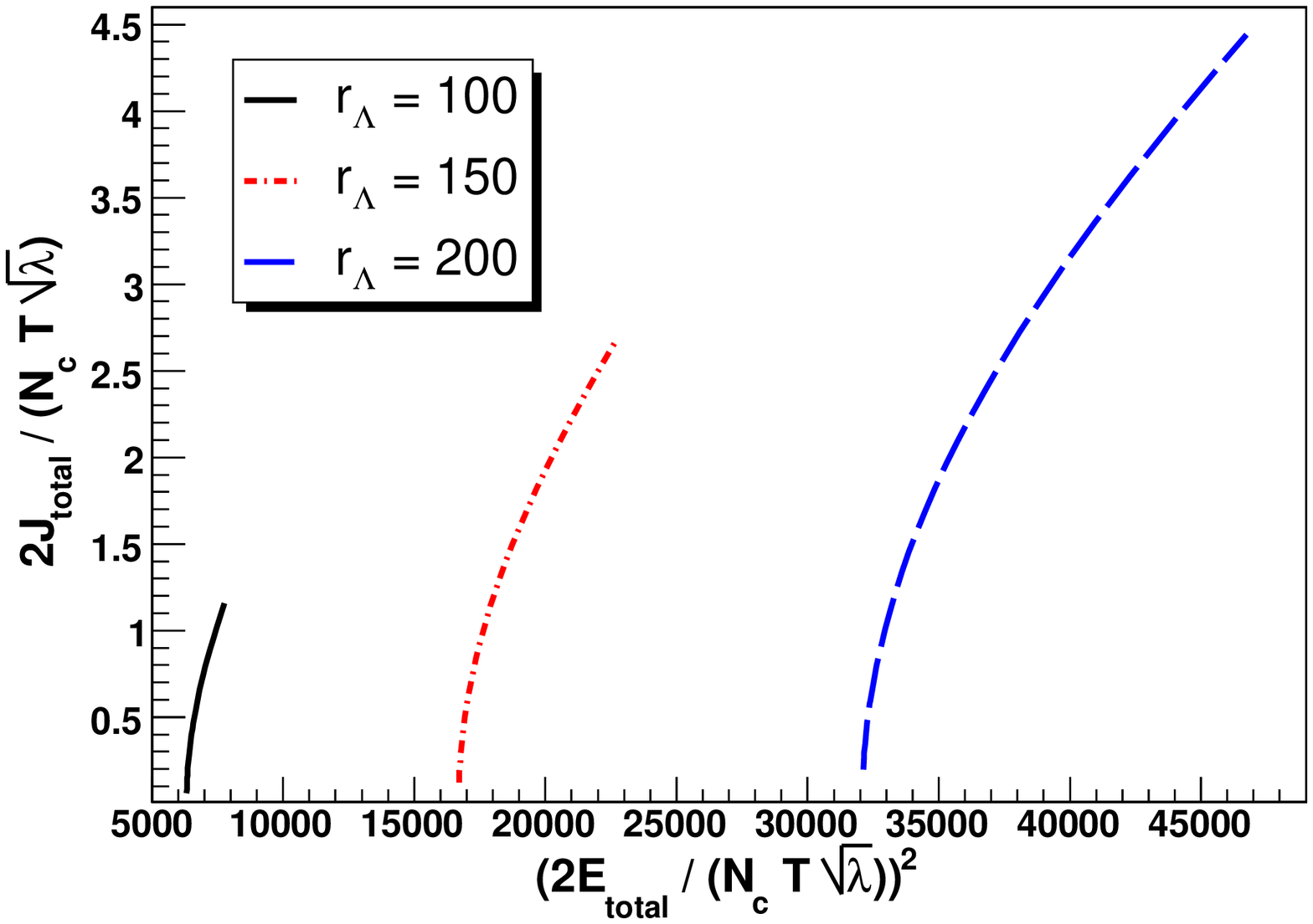}
  \caption{\small $J-E^2$ behavior of baryon with fixed
 $l_q$ on the boundary.
    }\label{fig:E_J_lq}
\end{minipage}
 \begin{minipage}[t]{0.48\linewidth}
  \centering
  \includegraphics*[width=1.0\columnwidth]{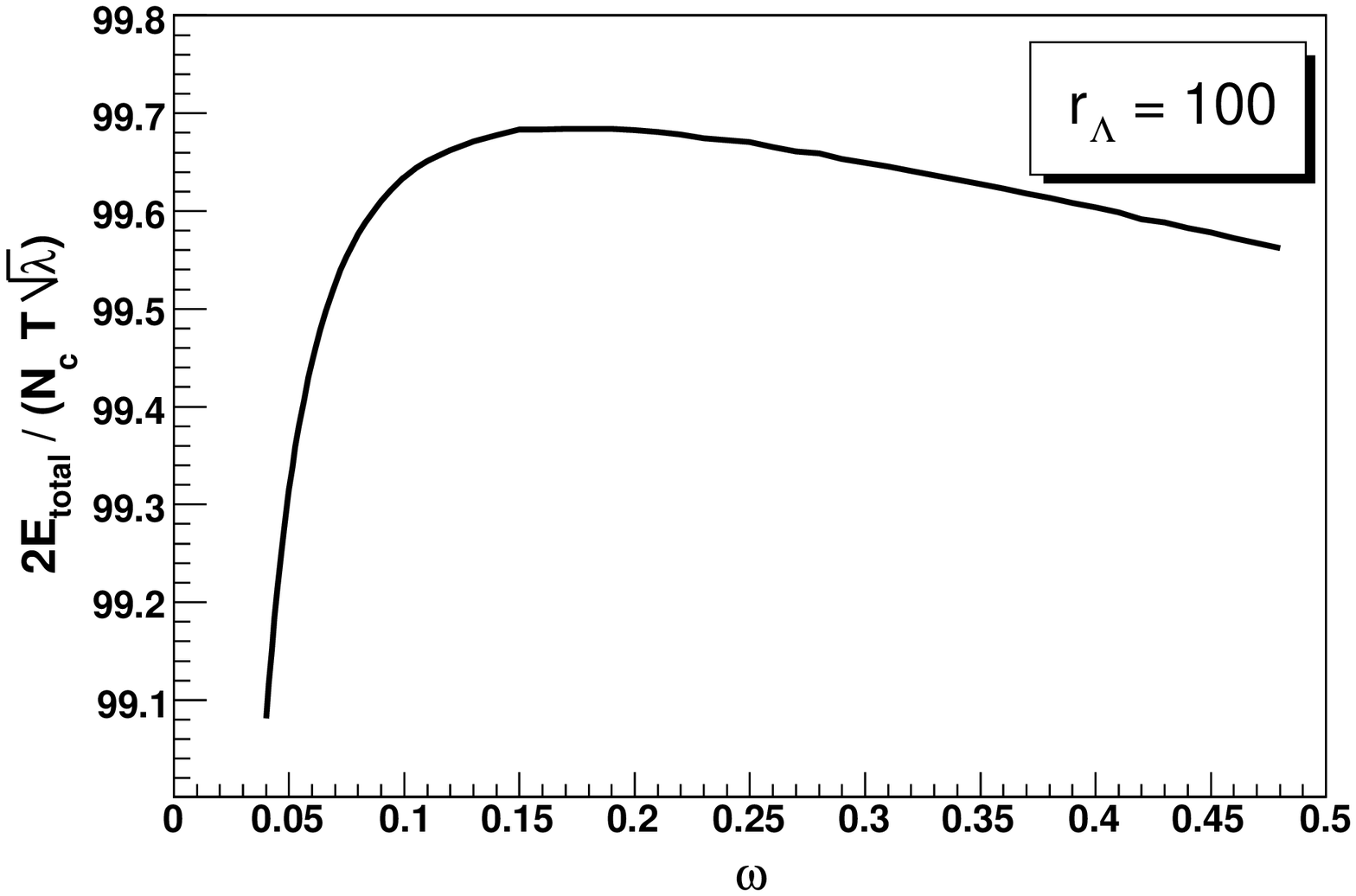}
  \caption{\small $\omega$ dependence of baryon energy
  with orthogonal boundary condition.
   }\label{fig:E_omega_ph}
 \end{minipage}
\begin{minipage}[t]{0.48\linewidth}
  \centering
  \includegraphics*[width=1.0\columnwidth]{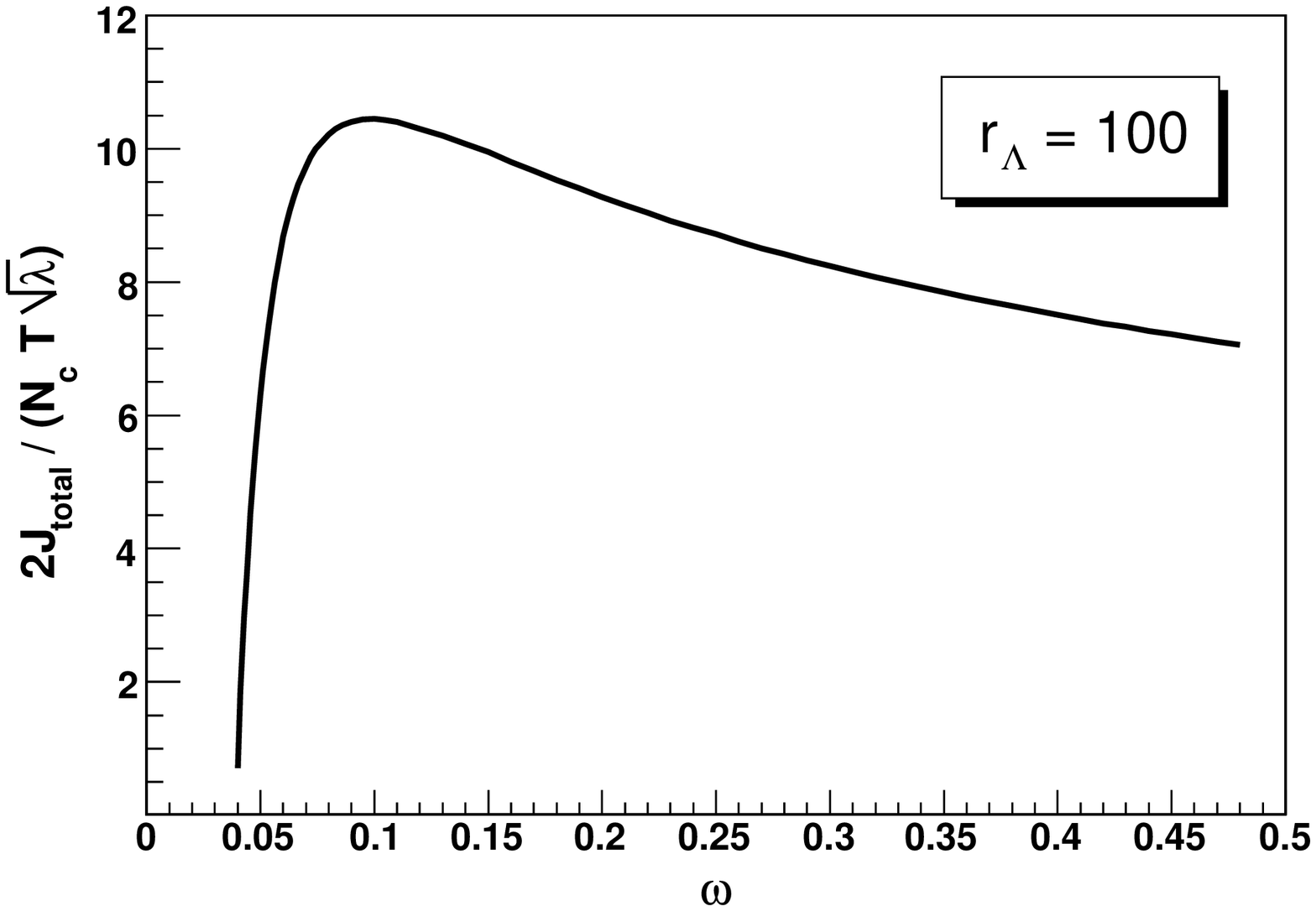}
  \caption{\small $\omega$ dependence of baryon $J$ charge with orthogonal boundary condition
    .}\label{fig:J_omega_ph}
 \end{minipage}
\end{figure}

\begin{figure}[t]
\begin{minipage}[t]{0.48\linewidth}
  \centering
  \includegraphics*[width=1.0\columnwidth]{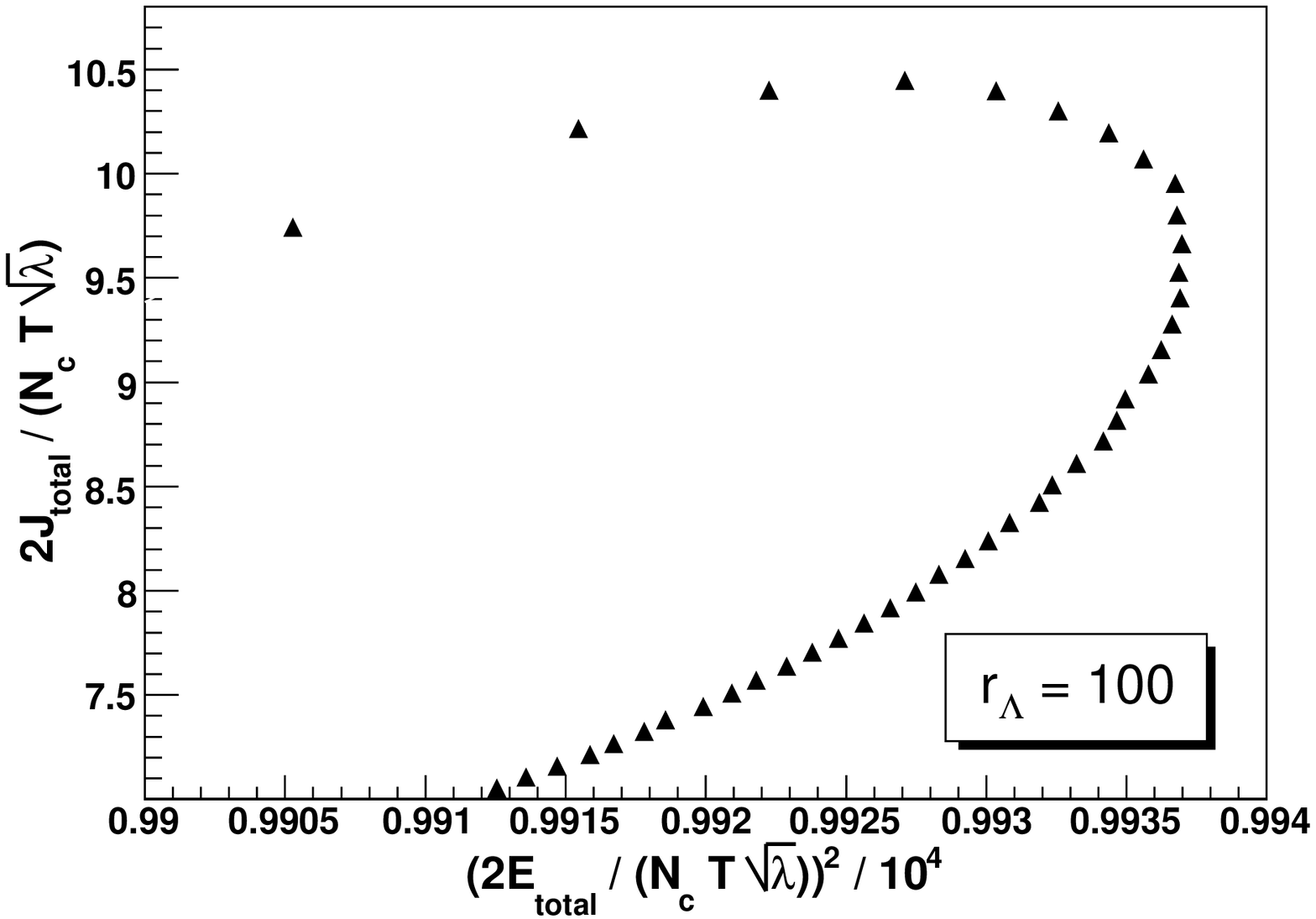}
  \caption{\small $J-E^2$ behavior of baryon with the orthogonal boundary
  condition.
  }\label{fig:E_J_single_ph}
  \end{minipage}
\begin{minipage}[t]{0.48\linewidth}
\centering
\includegraphics*[width=1.0\columnwidth]{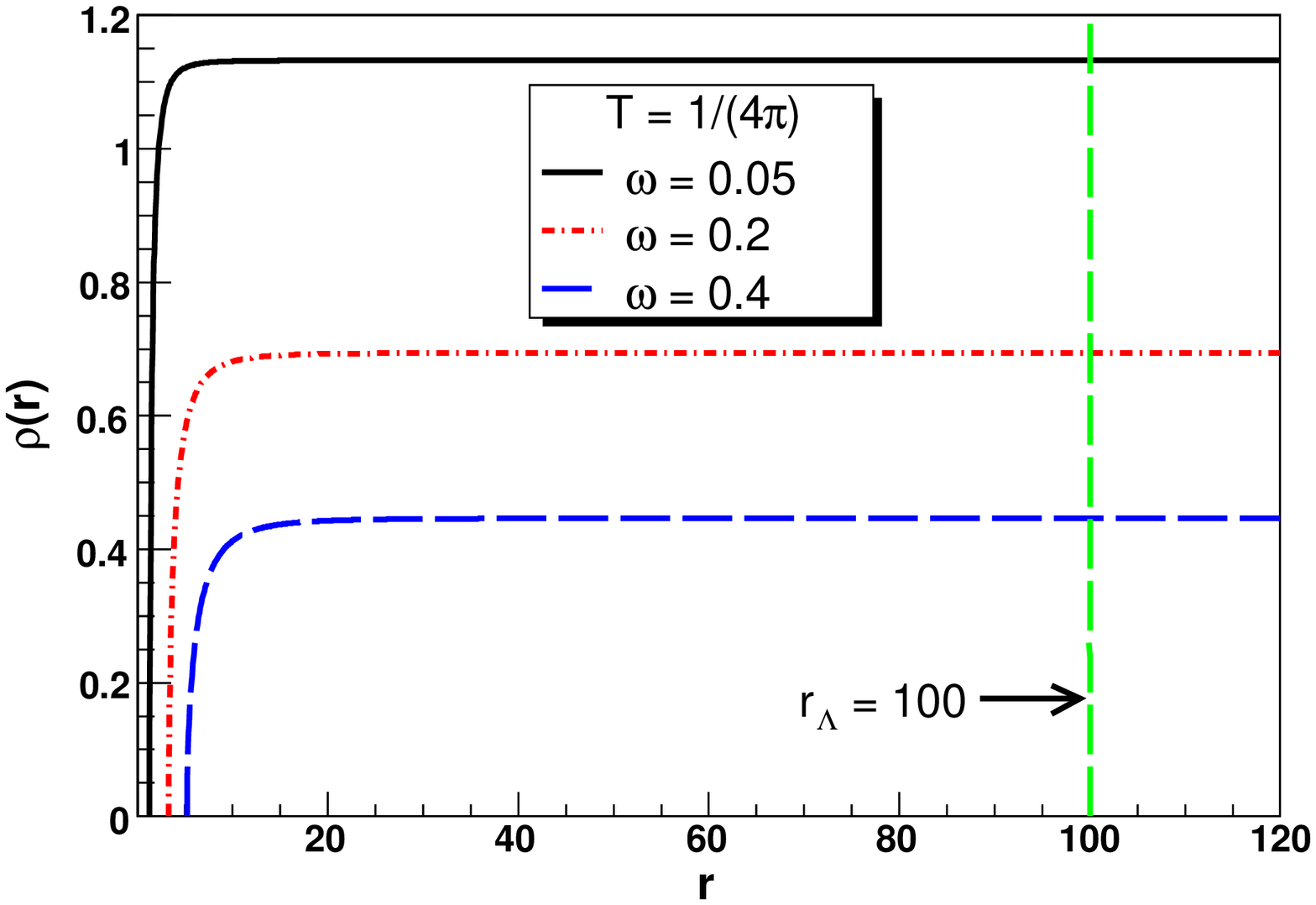}
\caption{\small Embedding function with orthogonal boundary
condition at different values of $\omega$. }\label{fig:E_J_rho}
\end{minipage}
 \end{figure}
The total energy and $J$ charge of the baryon can be defined from
the bulk strings and compact D5 brane \be\label{Etotal}
 E_{total}=N_cE_{string}+E_{brane},
\ee and \be\label{Jtotal}
 J_{total}=N_cJ_{string}+J_{brane},
\ee where $E_{string}\,,\;J_{string}$ are string energy and $J$
charge with $r_{\Lambda}$ cutoff.  Using the above equations
(\ref{jcharge})(\ref{stringenergy})(\ref{Ebrane})(\ref{Etotal})(\ref{Jtotal}),
finally we obtain \be \begin{split}
 E_{total}=&{N_c \ov 2\pi\alpha'}\int_{r_e}^{r_\Lambda}dr\frac{\left(\frac{1}{f(r)}+\frac{r^2}{R^2}\rho'^2(r)\right)f(r)}
{\sqrt{-\left(\frac{r^2}{R^2}\rho^2\omega^2-f(r)\right)
\left(\frac{1}{f(r)}+\frac{r^2}{R^2}\rho'^2(r)\right)}}+{{\VV(r_e)V_5}
\ov (2\pi)^5\alpha'^3}\;,
\end{split}
\ee and
 \be
\begin{split}
 J_{total}=&{N_c \ov 2\pi\alpha'}\int_{r_e}^{r_{\Lambda}}dr\frac{\left(\frac{1}{f(r)}+\frac{r^2}{R^2}\rho'^2(r)\right)(\frac{r^2}{R^2}\rho^2\omega)}
{\sqrt{-\left(\frac{r^2}{R^2}\rho^2\omega^2-f(r)\right)
\left(\frac{1}{f(r)}+\frac{r^2}{R^2}\rho'^2(r)\right)}}+J_{brane}\
\end{split}
\ee The energy of a hanging string is very large for the mass of
free quark, which corresponds to the straight string energy. To
define the interaction potential $E_I$, we need to subtract the
$N_c$ free quarks mass\footnote{At first, we subtract the rotating
$N_c$ free quarks mass $E_q={N_c \ov
2\pi\alpha'}\int_{r_0}^{r_{\Lambda}} dr
 ({1 \ov\sqrt{1-\rho^2(r_{\Lambda})\omega^2}})$ and at $\omega\neq0$ we see a smooth curve
along which $E$ and $l_q$ achieve their maximum values at different
points in Figure \ref{fig:EI-lq}(v1 of work~\cite{Li:2008py}). It
leads to some puzzles. We thank the editor for pointing out the
important question.}
 \be
 E_q={N_c \ov 2\pi\alpha'}\int_{r_0}^{r_{\Lambda}} dr\;,
 \ee then the interaction
potential can be given
\begin{equation}\label{EI}
E_I=N_cE_{string}-E_q+E_{brane} \;.
\end{equation}
We find the $l_q$ dependence of the $E_I$ in Figure \ref{fig:EI-lq}.
In this figure, the interaction potential also has a critical
behavior at the critical value of $l_q(l_q=L_c)$. For the given
$l_q$ we choose the low energy as the energy of stable baryon, and
the lower energy also corresponds to the larger $r_e$ in Figure
\ref{fig:lq-re2}.
We investigate $\omega$ dependence of the baryon energy and $J$
charge with three different conditions. The first one, we fix the
$r_e$. The configurations with different $\omega$ are shown in
Figure \ref{fig:rho-r}. We compute $\omega$ dependence of the energy
and charge and give the numerical result in figure
\ref{fig:E-omega},\ref{fig:J-omega}\footnote{We should note that in
Figure \ref{fig:E-omega}, the left part of each curve divided by the
green line corresponds to points with $r_e=3$ on different curves
which include the point ($r_e=3$) on their right arms in Figure
\ref{fig:lq-re2}. The right arms of curves in Figure
\ref{fig:lq-re2} are considered as configurations of real baryon
probes.(There are same arguments for Figure \ref{fig:J-omega} and
\ref{fig:regg}). }. Numerical results about $J_{total}-E_{total}^2$
relation are shown in Figure \ref{fig:regg}. In figure
\ref{fig:E-omega},\ref{fig:J-omega}, three different curves
correspond to three different cutoffs. We can see the $\omega$
dependence of $E_{total}$ and $J_{total}$. We find that as $\omega$
increases from zero, the energy and charge become larger at first
and then decrease. The curves have highest points both for the
energy and charge. We can explain this phenomena. Two main factors
of the variance of energy are angular velocity $\omega$ and the
rotating radius $\rho$. The former is dominant at first and the
latter is dominant when $\omega$ is big enough. The same thing
happens to the $J$ charge. We find that there are two branches for
the $J_{total}-E_{total}^2$ relation of baryons.

 The second one, we fix the quark separation $l_q$ on the boundary
 brane. It implies that we consider the baryons with almost the same size. We compute $\omega$ dependence of the energy and
 charge numerically in Figure \ref{fig:E_omega_lq}, \ref{fig:J_omega_lq}.
We give the $J-E^2$ behavior in Figure \ref{fig:E_J_lq}.

 At last, we choose a boundary condition which will be discussed as follow.
We want to analyze the boundary condition on the cutoff brane. In
section 2, we get the Nambu-Goto action after parameterizing the
string world sheet in spacetime:
\begin{equation}\
S_{string}=\frac {\TT}{2\pi \alpha'}\int_{r_e}^{r_{\Lambda}}dr
\sqrt{-\left(\frac{r^2}{R^2}\rho^2\omega^2-f(r)\right)
\left(\frac{1}{f(r)}+\frac{r^2}{R^2}\rho'^2(r)\right)}\;.
\end{equation}
Variation of the action gives the boundary term: \be
 \frac {\TT}{2\pi \alpha'}{\p\LL\ov\p\rho'}\delta\rho\biggr|_{r_e}^{r_\Lambda}
\ee In section 3, we calculate the screening length by considering
$\delta\rho(r_\Lambda)=0$. Because we think that the orbit of the
rotating string is exactly a close circle and the radius never
changes for the given $\omega$. But $\delta\rho(r_\Lambda)\neq0$
when we try to change $\omega$. We want to pick up some
configurations which can stand for the baryons, with same component
quarks but different $\omega$, and analyze $\omega$ dependence of
the energy and charge. So we choose the boundary condition
satisfying ${\p\LL\ov\p\rho'}=0\;$, from which we obtain

\be\label{lightspeed}
 \frac{1}{4} \:l_q^2 \: \omega^2=(1-{r_0^4 \ov
r_\Lambda^4})\ee

or

\be\label{orth}
 \rho'(r_\Lambda)=0.
\ee We see that the end point of string moves with light speed on
the cutoff brane with the first condition (\ref{lightspeed}). While
the string is orthogonal to the brane with the second condition
(\ref{orth}).

 We now want to give more details about the embedding of the string with angular
velocity $\omega$ and define properties of baryon with these two
boundary conditions. We want to look at the condition
(\ref{lightspeed}) first. Unfortunately after analyzing a certain
$\rho(r)$ with small $\omega$, we can not find a light speed point.
Then we look at the condition (\ref{orth}). From our numerical
result in the section 3, no matter what value of $r_e$ is given, the
strings are always orthogonal to the boundary plane in the
infinitely point in radial direction. However, for our cutoff
$r_\Lambda$, we can only choose a solution satisfying the orthogonal
condition in finite point $r_\Lambda$, which corresponds to the
highest point in Figure \ref{fig:rho-r}. Thus we can pick up the
string configurations satisfying this orthogonal boundary condition
as shown in Figure \ref{fig:E_J_rho}.

All the hanging strings are orthogonal to the cutoff brane. This
condition provide us a requirement to pick up one kind of
configurations with different spins\footnote{Here, the baryons with
orthogonal boundary condition are still unstable if we consider the
downward force on the boundary brane. However, this condition can
provide us a reasonable requirement for picking up configurations
with different $\omega$. The results in Figure \ref{fig:E_J_rho}$\;$
satisfy (\ref{fbc}), which implies that the configurations with this
condition do not have other external forces. That is natural and
very different from the two former conditions (fixed $r_e$ or
$l_q$).}. Thus we can pick up them and investigate $\omega$
dependence of $E_{total}$ and $J_{total}$. The corresponding
embedding functions are shown in Figure \ref{fig:E_J_rho}. The
boundary separation $\rho(r_\Lambda)$ becomes small when we increase
the $\omega$. Assuming $\omega\sim\rho(r_\Lambda)^n$. From Figure
\ref{fig:E_J_rho}, we can set $n=-1.5$. Then we consider a rough and
simplest rotating quark picture on the boundary brane. The force
balance condition in $\rho$ direction gives

\be\label{fbc}
 {\p E_I(\rho) \ov \p\rho}=m_q\omega^2\rho
\ee where the right part is the centrifugal force. Using
$\omega\sim\rho(r_\Lambda)^{-1.5}$, we obtain $E_I\sim\rho^{-1}$.
Here, we just ignore the variance of the function $E_I(\rho)$ when
$\omega$ increases from zero. Let us turn to Figure \ref{fig:EI-lq}
and look the numerical result. If we fit the curve with $\omega=0.5$
by using \be
 {2E_I \ov N_cT\sqrt{\lambda}}={al_q ^c+b},(\rho(r_\Lambda)=l_q)
\ee we obtain
\be
 a=-0.060, b=0.762, c=-1.018
\ee while we get \be a=-0.637, b=0.949, c=-1.008 \ee for the curve
with $\omega=0$. The values of $c$ satisfy $E_I\sim\rho^{-1}$.

We compute the $\omega$ dependence of the energy and charge as in
Figure \ref{fig:E_omega_ph}, \ref{fig:J_omega_ph}. And, we compute
$J-E^2$ behavior numerically in Figure \ref{fig:E_J_single_ph}. The
curve has two branches and we fit them by using the linear function
respectively. We choose the fit function as \be
 {2J \ov N_cT\sqrt{\lambda}}=a({2E\ov N_cT\sqrt{\lambda}})^2+b,
\ee then we get \be a=601.606, b=-586.255.\ee with small $\omega$
corresponding to the upper branch and \be a=652.6, b=-639.9 \ee with
large $\omega$ corresponding to the lower branch. The value of $a$
can be used to detemine the Regge slope, with the coefficients
before $E$ and $J$.

\section{Discussion and Conclusion}
  In this paper, we calculate the $\omega$ dependence of the baryon
screening length in hot plasma. We use the strongly coupled SYM
gauge theory to describe the quark gluon plasma and consider baryon
as a probe. By considering a simple model of baryon in the AdS/CFT
frame, we investigate the signatures of baryon probe through the
bulk calculation, which are useful to properties of probe of in
strongly coupled gauge theory. In particular, we consider baryons
with high spin using the rotating strings for the first time, and
obtain the $\omega$ dependence of the embedding function of hanging
strings and screening length. The relation between screening length
and spin may be an important experiment signal for the QGP. The
energy and charge have been defined for the high spin baryon in our
paper. We investigate the $J-E^2$ behavior of baryons, and give the
numerical results in three different conditions in our paper. Among
them, we argue that these solutions with orthogonal boundary
condition are the best candidates for baryons with same component
quarks but different spins. The most important is that we obtain two
branches of $J-E^2$ behavior. The slope is large for large $\omega$
as shown in Figure \ref{fig:E_J_single_ph}. We ignore the
$J_{brane}$, which means there is no intrinsic spin of D5 brane.
These behaviors are probably most important properties of baryons in
strongly coupled plasma.

However, in real relativistic heavy ion collisions quark gluon
plasma is different from $\mathcal {N}=4$ SYM. To achieve the goal
of understanding phenomena in relativistic heavy ion collisions,
conformal invariance is broken via an one-parameter deformation of
the AdS black hole dual to the hot $\mathcal{N}=4$ SYM plasma, and
robustness with respect to the introduction of nonconformality of
five observables of strongly coupled plasmas that have been
calculated in $\mathcal{N}=4$ SYM theory at nonzero temperature has
been investigated \cite{Liu:2008RS,Caceres:2006ta}. The result shows
that, in a toy model, the jet quenching parameter and screening
length are affected by the nonconformality at the $20\sim30\%$ level
or less, but the drag and diffusion coefficients for a slowly moving
heavy quark are modified by as much as $80\%$.

On the other hand, considering the QGP be nonsupersymmetric, we can
use other realistic geometry backgrounds. We can compactify a space
dimension and give antiperiodic boundary condition for the fermions
to break supersymmetry completely. In the baryon model in our paper,
the compact brane sits at the same point in the AdS space and has a
simple action. Actually the compact brane can be always very heavy
and has a nontrivial trajectory in the AdS. For example, in general
case the trajectory of the compact brane in the AdS space (including
time direction) can be a plane, but not a line.

If we consider different compact D branes, the embedding function of
the brane in the bulk can be determined by the DBI action. In this
case, the compact brane can be enlongated to mimic a string, so the
lengths of the string tend to be
zero~\cite{Seo:2008,Callan:1998iq,Imamura:1998gk,Ghoroku:2008tg}.
Properties of baryon will be effected by the brane configuration.
Along this way, we can get more interesting properties of the dual
baryon model in AdS/CFT and get more informations of the baryon
probe in hot plasma.

After this work was finished, there appears very recent work
investigating the properties of baryon in strongly coupled gauge
field~\cite{Ghoroku:2008na}.

\section*{Acknowledgements}
Yang Zhou acknowledges helpful discussions with Tower Wang, Gang
Yang, Yushu Song, Chaojun Feng, Yi Wang, Huanxiong Yang and thanks
Xian Gao for kind help. Push Pu would like to thank Jian Deng for
numerical discussions and Qun Wang for kind help. We thank
A.~Guijosa for pointing out more references very clearly. We also
are very grateful to the editor for kind comments and very important
suggestions.

\end{document}